\documentclass[review,times,sort&compress,number]{elsarticle}
\usepackage{lineno}
\usepackage{hyperref}
\usepackage{xurl}
\usepackage{amssymb}
\usepackage{subcaption} 
\usepackage{booktabs, multirow} 
\usepackage{array}
\usepackage{amsmath} 
\usepackage{multirow}
\usepackage{makecell}
\usepackage{xcolor}

\newcommand{ \sect }[1]{Sec.~\ref{#1}}
\newcommand{ \eq }[1]{Eq.~(\ref{#1})}
\newcommand{ \fig }[1]{Fig.~\ref{#1}}
\newcommand{ \tab }[1]{Table~\ref{#1}}

\newcommand{ \mm }{\ensuremath{\rm mm}}
\newcommand{ \mmp }[1]{\ensuremath{\rm mm^{#1}}}
\newcommand{ \nsec }{\ensuremath{\rm ns}}
\newcommand{ \usec }{\ensuremath{\rm \mu s}}
\newcommand{ \MeVc }{\ensuremath{{\rm MeV}/c}}

\newcommand{ \PhaseAlpha }{Phase-$\alpha$}
\newcommand{ \piMone }{$\rm \pi M1$}
\newcommand{ \mucap }{muon capture}
\newcommand{ \T }[1]{\ensuremath{\rm T_{#1}}}
\newcommand{ \Tz }[1]{\ensuremath{\rm T_0^{{#1}0}}}
\newcommand{ \eff }[1]{\ensuremath{\epsilon_{\rm #1}}}
\newcommand{ \pur }[1]{\ensuremath{\eta_{\rm #1}}}
\newcommand{ \accep }[1]{\ensuremath{A_{\rm #1}}}
\newcommand{ \N }[1]{\ensuremath{N_{\rm #1}}}
\newcommand{ \R }[1]{\ensuremath{R_{\rm #1}}}
\newcommand{ \ltime }[1]{\ensuremath{\tau_{\rm #1}}}

\begin{document}
\begin{frontmatter}

\title{Development of the Range Counter for the COMET \PhaseAlpha{} Experiment}

%%%%%%%%%%%%%%%%%%%%%%%%%%%%%%%%%%%%%%%%%%%%%%%%%%%%%%%%%%%%%
% Authors
%%%%%%%%%%%%%%%%%%%%%%%%%%%%%%%%%%%%%%%%%%%%%%%%%%%%%%%%%%%%%
\author[KEK,ICL]{Kou~Oishi\corref{cor1}}
\ead{kou.oishi@kek.jp}
\cortext[cor1]{Corresponding author}

\author[Osaka]{Masaharu~Aoki}
\author[Osaka]{Shion~Kuribayashi}
\author[Osaka]{Shunya~Ueda}
\author[Osaka]{Kazuki~Ueno}
\author[Clermont]{Nicolas~Chadeau}
\author[Clermont]{Thomas~Clouvel}
\author[ICL]{Yuki~Fujii}
\author[KEK]{Yoshinori~Fukao}
\author[Sokendai]{Masaaki~Higashide}
\author[KEK]{Youichi~Igarashi}
\author[KEK]{Satoshi~Mihara}
\author[KEK]{Hajime~Nishiguchi}
\author[Sokendai]{Kenya~Okabe}
\author[ICL]{Yoshi~Uchida}

%%%%%%%%%%%%%%%%%%%%%%%%%%%%%%%%%%%%%%%%%%%%%%%%%%%%%%%%%%%%%
% Affiliation
%%%%%%%%%%%%%%%%%%%%%%%%%%%%%%%%%%%%%%%%%%%%%%%%%%%%%%%%%%%%%
\affiliation[KEK]{
    organization={High Energy Accelerator Research Organization},
    addressline={1-1 Oho}, 
    city={Tsukuba, Ibaraki},
    postcode={305-0801}, 
    country={Japan}
}
        
\affiliation[ICL]{
    organization={Imperial College London},
    addressline={Exhibition Road}, 
    city={London},
    postcode={SW7 2BX}, 
    country={United Kingdom}
}

\affiliation[Osaka]{
    organization={The University of Osaka},
    addressline={1-1 Machikaneyama}, 
    city={Toyonaka, Osaka},
    postcode={560-0043}, 
    country={Japan}
}

\affiliation[Clermont]{
    organization={Laboratoire de Physique de Clermont (LPC), Université Clermont Auvergne},
    addressline={4 Avenue Blaise Pascal}, 
    city={Clermont-Ferrand},
    postcode={63178}, 
    country={France}
}

\affiliation[Sokendai]{
    organization={SOKENDAI},
    addressline={1-1 Oho}, 
    city={Tsukuba, Ibaraki},
    postcode={305-0801}, 
    country={Japan}
}

%%%%%%%%%%%%%%%%%%%%%%%%%%%%%%%%%%%%%%%%%%%%%%%%%%%%%%%%%%%%%
% Abstract
%%%%%%%%%%%%%%%%%%%%%%%%%%%%%%%%%%%%%%%%%%%%%%%%%%%%%%%%%%%%%
\begin{abstract}
The COMET \PhaseAlpha{} experiment aims to evaluate the novel muon transport beamline for the muon-to-electron conversion search at J-PARC, Japan.  
A dedicated Range Counter (RC) was developed to measure the momentum spectrum of transported negative muons with momenta of 30--100~\MeVc.  
The RC consists of graphite momentum degraders, a muon absorber, and plastic scintillation counters (\T0, \T1, and \T2) to detect decay-in-orbit (DIO) electrons from stopped muons.  
The number of muons stopped in the absorber is reconstructed from the decay time distribution.
A copper absorber was selected due to the short lifetime of muonic atoms in copper, which enhances signal separation.
The counters' performance was evaluated experimentally.  
The \T0 Counter, made of a $200\times 200\times 0.5~\mmp3$ scintillator plate, achieved a muon-trigger efficiency exceeding 99.9\%.  
The \T1 and \T2 Counters also demonstrated high electron-detection efficiencies of $>99\%$.  
Based on these results, simulation studies estimate the acceptance for reconstructing the number of DIO electrons from the absorber to be approximately 47\% with a corresponding signal purity of 60\% against muon capture-induced backgrounds.
\end{abstract}


\end{frontmatter}

%%%%%%%%%%%%%%%%%%%%%%%%%%%%%%%%%%%%%%%%%%%%%%%%%%%%%%%%%%%%%
% Main Text
%%%%%%%%%%%%%%%%%%%%%%%%%%%%%%%%%%%%%%%%%%%%%%%%%%%%%%%%%%%%%

%%%%%%%%%%%%%%%%%%%%%%%%%%%%%%%%%%%%%%%%%%%%%%%%%%%%%%%%%%%%%
% Introduction
%%%%%%%%%%%%%%%%%%%%%%%%%%%%%%%%%%%%%%%%%%%%%%%%%%%%%%%%%%%%%
\section{Introduction}
\label{sec:Introduction}

The COMET experiment searches for the muon-to-electron conversion process in aluminium atoms, $\mu^- + {\rm Al} \rightarrow e^- + {\rm Al}$~\cite{COMET}.
This process violates charged lepton flavour conservation and is therefore extremely suppressed in the Standard Model (SM) with neutrino oscillations, yielding a branching ratio well below $\mathcal{O}(10^{-50})$.
In contrast, it can be significantly enhanced in several theories beyond the SM (BSM) up to $\mathcal{O}(10^{-15})$.
Its experimental observation would thus provide direct evidence of BSM physics.

The experiment is conducted in two phases, Phase-I and Phase-II, at the Japan Proton Accelerator Research Complex (J-PARC) in Japan.
To achieve a single-event sensitivity of $\mathcal{O}(10^{-15})$ in Phase-I, a large number of negative muons are produced by irradiating a graphite target with the high-intensity proton beam of J-PARC and transported to the experimental area. 

A dedicated muon transport beamline is under construction, with the $90^{\circ}$-curved Muon Transport Solenoid (TS)~\cite{TS.1, TS.2} playing a crucial role in transporting slow negative muons with a momentum of 30--50~\MeVc{}, while simultaneously filtering out high-momentum or positively charged particles causing backgrounds.
It exploits the fact that charged particles flying along the curved solenoidal field experience vertical drift relative to the transport direction.
The drifting direction and magnitude depend on the charge polarity and momentum.  
Dipole magnetic fields compensate for this drift, ensuring that only negative muons within the desired momentum range reach the detectors, with beam collimators applied as needed.
The first beamline-commissioning project, \PhaseAlpha{}, was planned to verify the TS performance and investigate the transported muon beam in the COMET experimental area.

\begin{figure}[htb]
    \centering
    \includegraphics[width=\columnwidth]{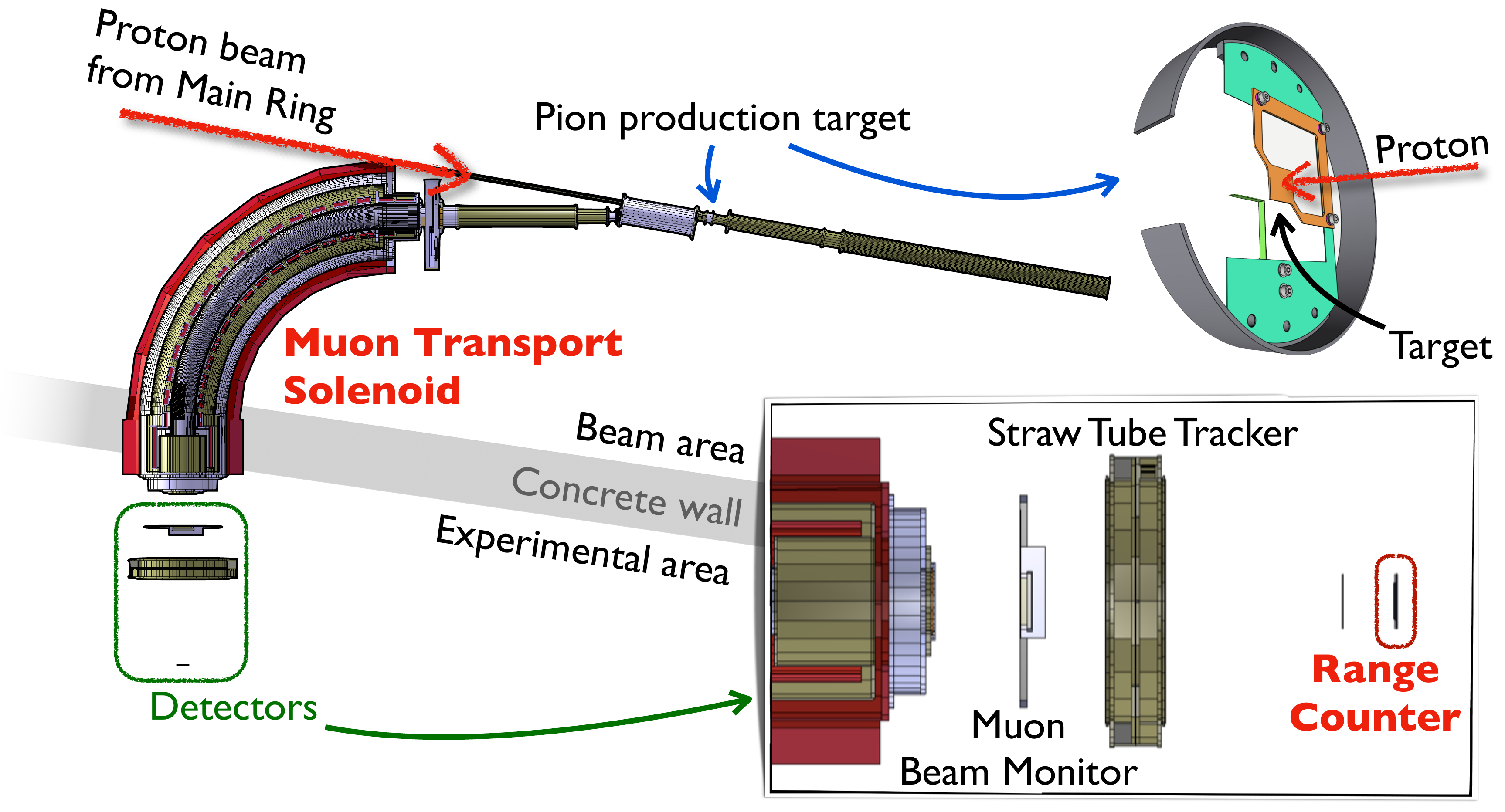}
    \caption{
        Experimental setup of \PhaseAlpha.
        The proton beam from the J-PARC Main Ring is injected into the pion production target, producing secondary pions and muons.  
        Low-momentum muons are transported through the Muon Transport Solenoid to the experimental area, where they are detected by the Muon Beam Monitor, Straw Tube Tracker, and Range Counter.  
    }
    \label{fig:BeamlineSetup}
\end{figure}

\fig{fig:BeamlineSetup} shows the \PhaseAlpha{} experimental setup.  
The J-PARC Main Ring synchrotron accelerator provides a bunched high-intensity 8~GeV proton beam to the COMET proton beamline.  
The beam protons are injected into the pion production target, which is a thin carbon-composite sheet with dimensions of $20\times 20\times 1.1~\mmp3$.
Backward-emitted secondary pions and muons enter the TS, where low-momentum muons are transported to the experimental area.  

\PhaseAlpha{} includes three detectors: the Muon Beam Monitor (MBM)~\cite{MBM}, the Straw Tube Tracker (STT)~\cite{StrawTrk}, and the Range Counter (RC).  
The MBM and STT\footnote{
    In COMET Phase-I, the STT will comprise five hodoscope modules for tracking, whereas \PhaseAlpha{} employs a single module for beam profiling.  
}  
are hodoscope detectors that measure a two-dimensional profile of the secondary beam.
The MBM consists of plastic fibres with a cross section of $1\times 1~\mmp2$.
The STT consists of gaseous drift tubes, called straw tubes, with a diameter of 9.8~mm.  

The RC is the primary detector in \PhaseAlpha{} for measuring the momentum spectrum of the transported muons in the range of 30--100~\MeVc.
It was designed with an ultra-thin 0.5~mm plastic scintillator, covering a large area to capture as many muons as possible in a single measurement.  
This paper describes its design and performance.

%%%%%%%%%%%%%%%%%%%%%%%%%%%%%%%%%%%%%%%%%%%%%%%%%%%%%%%%%%%%%
% Range Counter
%%%%%%%%%%%%%%%%%%%%%%%%%%%%%%%%%%%%%%%%%%%%%%%%%%%%%%%%%%%%%
\section{The Range Counter}
\label{sec:Design}

\begin{figure}
    \centering
    \includegraphics[width=\columnwidth]{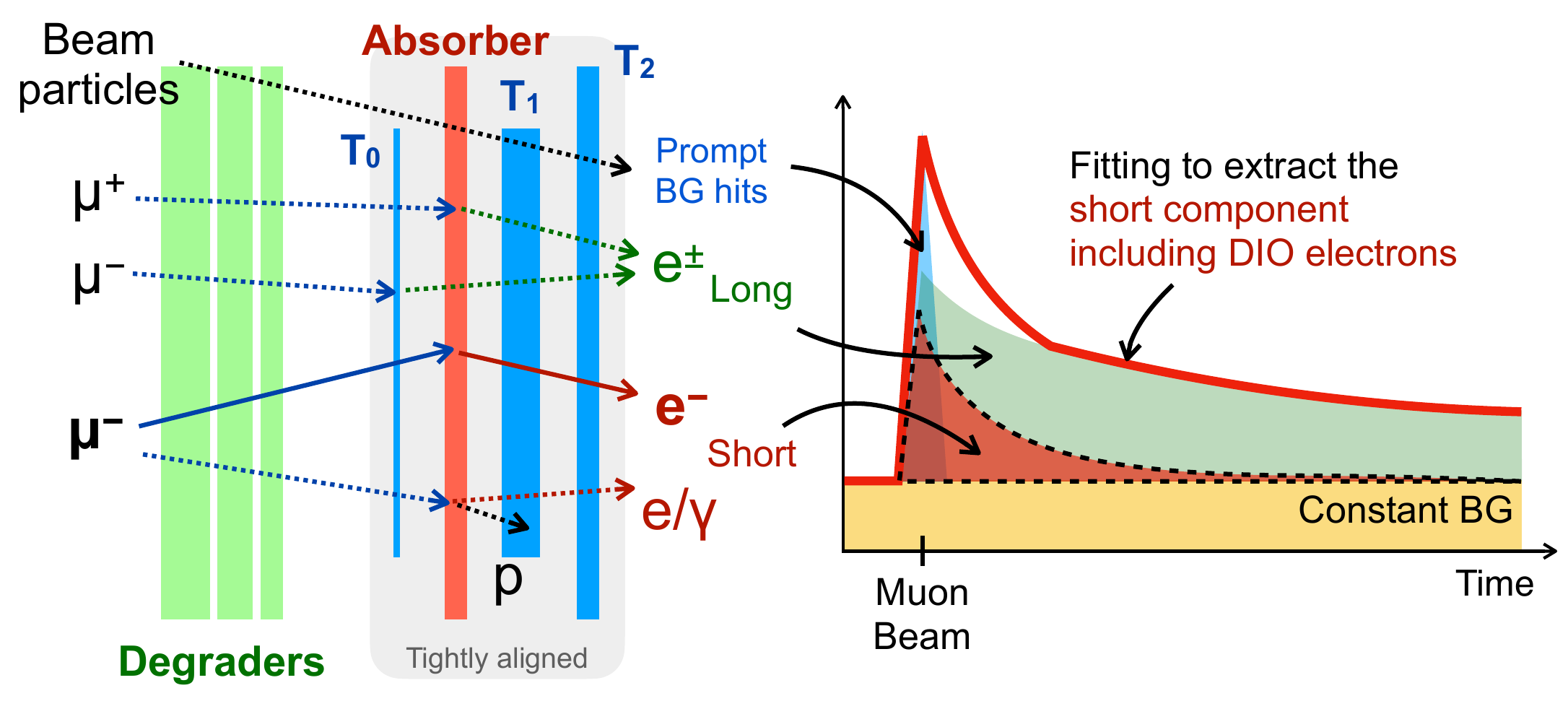}
    \caption{
        Schematic view of the Range Counter and its measurement principle.
        Muons slowed down by the degraders stop in the absorber.
        The \T0 Counter triggers the muons and measures their timing, and the \T1 and \T2 Counters then detect outgoing particles to reconstruct a time distribution.
        The components in the gray region are tightly aligned without gaps to maximise geometrical acceptance.
        An appropriate function is fitted to the distribution to count the short-lifetime component (red), which includes the signal DIO electrons, separately from the backgrounds (green, blue, and yellow).
    }
    \label{fig:RangeCounter}
\end{figure}

The RC is an assembly of momentum-degrading plates, a muon-absorbing plate, and three timing counters named the \T0, \T1, and \T2 Counters.  
Its objective is to statistically count the number of beam muons trapped in its absorber for several momentum ranges.
\fig{fig:RangeCounter} illustrates the measurement principle with a lateral-view schematics.
The \T0 Counter triggers data acquisition (DAQ) by detecting beam particles and determines the time reference for the event.
Incoming negative muons are slowed down by the degraders and stop in the absorber, where they form muonic atoms.
The degraders regulate the momentum range of the muons stopping within the absorber by their total thickness.  

Part of the stopped muons undergo decay-in-orbit (DIO), emitting electrons that are detected by the \T1 and \T2 Counters.  
We require DIO electrons to produce simultaneous hits in them with sufficient energy deposits, as well as all the timing counters and absorber must be installed close to each other for a better geometrical acceptance.
The \T1 and \T2 Counters are used to construct distributions of the muons’ decay time, defined as $T_{\rm decay} = (T_1 + T_2)/2 - T_0$, where $T_i$ is the timing measured by the corresponding \T{i} Counter.
Averaging the timings from both \T1 and \T2 Counters aims to improve the time resolution.

The lifetime of a captured negative muon is nucleus dependent, serving as a distinctive characteristic of DIO electrons.
The timing counters dominate the RC volume, and their materials are typically carbon rich; the muon lifetime in them is $\sim 2~\usec$, close to that of free muons.
To differentiate DIO electrons occurring in the absorber, the absorber must be made of a material with heavy atoms.
Then, the muonic atoms in it effectively have a shorter lifetime, $\ltime{short}$, and form a noticeable part of the distribution, referred to in this paper as the short component compared to the other long component with a long lifetime.
Consequently, the degraders also should be composed of lightweight materials.

% Total DIO electrons
The total number of short component detected by the \T1 and \T2 Counters, $\N{short}$, is reconstructed by fitting the following function to the observed $T_{\rm decay}$ distribution.
\begin{equation}
    F(T_{\rm decay}) = \frac{\N{short}}{\ltime{short}} \exp\left(-\frac{T_{\rm decay}}{\ltime{short}}\right) + {\rm BG\ terms},
    \label{eq:Tdecay}
\end{equation}
where the background (BG) terms contain the long component and constant distributions.

% The reason of double counters
The unavoidable BGs come from nuclear muon capture, which has the same lifetime as DIO: its secondary protons, $\gamma$-rays, and neutrons take up some of the short component.
To mitigate these, the \T1 Counter needs to be thick to prevent a significant fraction of protons from reaching the \T2 Counter.
In addition, appropriate thresholds are applied to their energy deposits in the counters to cut BG hits.
The \T1 and \T2 Counters are required to have a high detection efficiency for DIO electrons, $\eff{DIO}$, at the energy thresholds to apply.

% Other backgrounds
Other background (BG) sources include beam-induced particles contaminating the distribution, as well as environmental BGs with a constant hit rate.
The beam can contain positive muons; however, they only contribute to the long-lifetime component.
On the other hand, it is highly challenging to eliminate accidental hits around the time reference, which are induced by the bunched beam injection.
These prompt BGs must be taken into account both in the reconstruction process and in the material selection for the absorber, as discussed later.

The number of negative muons stopping in the absorber plate, \N{stop}, is reconstructed from \N{short} by 
\begin{equation}
    \N{stop} = \frac{\N{short}\ \pur{DIO}}{\eff{trig}\ \R{DIO}\ \accep{DIO}},
    \label{eq:Nstop}
\end{equation}
where \eff{trig} is the muon-trigger efficiency of the \T0 Counter, and \R{DIO} is the fraction that the muonic atoms induce DIO in the absorber.
\pur{DIO} represents the purity of the DIO contribution in \N{short} after applying the energy deposit cut.
\accep{DIO} includes both the geometrical acceptance of the \T1 and \T2 Counters for signal DIO electrons from the absorber and the analytical acceptance arising from the energy cut.

The geometrical acceptance can ideally reach 50\% when the downstream side of the absorber is fully covered.
These two factors are correlated, as DIO electrons passing near the counter edges may fail to deposit sufficient energy, which also affects \pur{DIO}.

\pur{DIO} and \accep{DIO} are formulated as
\begin{equation}
    \pur{DIO} = \frac{\N{DIO}^{\rm cut}}{\N{DIO}^{\rm cut} + \N{cap}^{\rm cut}}, \quad 
    \accep{DIO} = \frac{\N{DIO}^{\rm cut}}{\N{DIO}^{\rm all}}, 
    \label{eq:DIO}
\end{equation}
where $\N{DIO}^{\rm all}$ is the number of DIO electrons from the absorber, and $\N{DIO}^{\rm cut}$ denotes the number of those depositing energies exceeding the threshold.
$\N{cap}^{\rm cut}$ is the number of \mucap-induced hits that pass the energy cut and still remain within the short component. 

All the parameters in \eq{eq:Nstop} reflect the RC performance and hence must be sufficiently high.
We first estimated \R{DIO}, \pur{DIO}, and \accep{DIO} using literature values and simulation studies for the designed setup explained in \sect{sec:Design}.
In addition, we conducted an experiment to evaluate \eff{trig} of the \T0 Counter and \eff{DIO} of the \T1 and \T2 Counters, as described in \sect{sec:Performance}.

%%%%%%%%%%%%%%%%%%%%%%%%%%%%%%%%%%%%%%%%%%%%%%%%%%%%%%%%%%%%%
% Design
%%%%%%%%%%%%%%%%%%%%%%%%%%%%%%%%%%%%%%%%%%%%%%%%%%%%%%%%%%%%%
\section{Design}
\label{sec:Design}

Every timing counter consists of a wide, square plastic scintillator plate.
Eljen Technology EJ-212 is used for the \T0 Counter, while EJ-200 is used for the \T1 and \T2 Counters.
Although the difference between the two is minor for our purposes, EJ-212 is recommended for thin counters such as the \T0 Counter.
Light guides are essential both for guiding scintillation photons into the photo-sensors and for structurally supporting their coupling to the wide plate.
Fine-mesh photomultiplier tubes (PMTs), Hamamatsu H6154, are employed as the photo-sensors.
Each scintillator plate is equipped with one or two sets of an acrylic light guide and a PMT.
As described below, the \T0 Counter in particular requires two sets, installed on both sides.

As the entire RC assembly must remain compact to allow two-dimensional movement transverse to the beam direction within the limited detector space, the light guides are curved by $90^{\circ}$.
They are bonded to the scintillator plate using optical cement, Eljen~EJ-500, while the PMT is fixed to the light guide with optical grease and black vinyl light-shielding tape.
Each counter assembly is wrapped in aluminised Mylar and black sheets for light reflection and shielding.

%%%%%%%%%%%%%%%%%%%%%%%%%%%%%%%%%%%%%%%%%%%%%%%%%%%%%%%%%%%%%%%%%%%%%%%%%%%%%%%%%%%%%%%%%%%
% Degrader
%%%%%%%%%%%%%%%%%%%%%%%%%%%%%%%%%%%%%%%%%%%%%%%%%%%%%%%%%%%%%%%%%%%%%%%%%%%%%%%%%%%%%%%%%%%
\subsection{Degrader}

% Degrader
The degrader should cover all incoming particles from the beamline and be much lighter than the absorber material not to generate DIO electrons with a short lifetime.
We chose graphite plates with a cross section of $300\times 300~\mmp2$.
Graphite is an elementally pure form of carbon so that uncertainties in momentum control can be minimised, and it resists deformation due to heat and radiation.

The target muon momentum range in \PhaseAlpha{} is up to $\sim 100~\MeVc$, for which the degrader needs to have a thickness of 32~mm in total.
We prepared graphite plates with thicknesses of 4, 8, 16, and 32~mm.
Combinations of several plates out of those control the momentum range of stopping muons.

%%%%%%%%%%%%%%%%%%%%%%%%%%%%%%%%%%%%%%%%%%%%%%%%%%%%%%%%%%%%%%%%%%%%%%%%%%%%%%%%%%%%%%%%%%%
% T0 Counter
%%%%%%%%%%%%%%%%%%%%%%%%%%%%%%%%%%%%%%%%%%%%%%%%%%%%%%%%%%%%%%%%%%%%%%%%%%%%%%%%%%%%%%%%%%%
\subsection{$T_0$ Counter}
\label{sec:Design.T0}

The \T0 Counter is the most critical component and must be both thin and large to detect as many low-momentum muons as possible.  
A thickness of 0.5~mm was chosen to ensure the minimum required mechanical strength. 
This also allows measurement of negative muons down to approximately 30~\MeVc{}.
It is equipped with dual readout ports on the two lateral sides to compensate for its expected low light collection efficiency.

A preliminary study was conducted to determine the appropriate size of the \T0 Counter.  
This study investigated the light yield from a scintillator plate with dimensions of $300 \times 300 \times 0.5~\mmp3$.  
A Hamamatsu H7195 PMT was directly coupled to the centre of one side without a light guide to measure scintillation photons.
The entire setup was placed in a dark box.  
The plate was irradiated with collimated $\beta$-rays from a $\rm {}^{90}Sr$ source along a straight line from the PMT to examine light attenuation within the scintillator.
The detected photoelectron (p.e.) yield decreased by approximately 50\% as the irradiation point moved from 10~cm to 15~cm away from the PMT.

These results indicate that, given the realistic requirement to collect photons via light guides in the designed setup, a 30~cm-square counter would not provide sufficient light yield.
This concern led us to adopt a 20~cm-square scintillator as the baseline design, denoted in this paper as the \Tz2 Counter.  
In parallel, we also produced a 30~cm-square type, the \Tz3 Counter, and compared its performance with the \Tz2 Counter in terms of \eff{trig} to experimentally validate the design, as shown in \sect{sec:Performance}.

\begin{figure}[tb]
    \centering
    \includegraphics[height=0.2\textheight]{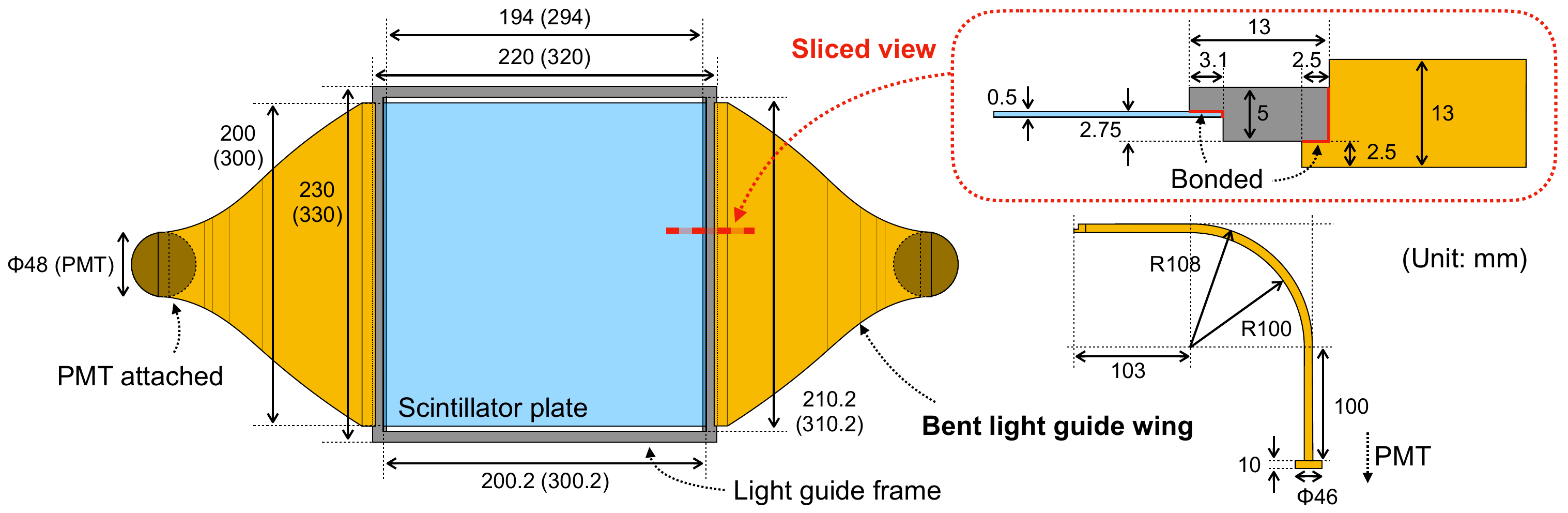}
    \caption{
        Schematics of the \T0 Counter using a 20~cm-square scintillator plate.  
        Values in parentheses correspond to the 30~cm-square type.  
        The main scintillator plate is supported by the light guide frame to ensure mechanical stability, accommodating the thicker and heavier bent light guides on the both sides.  
    }
    \label{fig:T0Design}
\end{figure}

\fig{fig:T0Design} is the schematics of the \T0 Counter.  
A hollow square light guide frame holds the fragile scintillator firmly at its centre, supporting the heavier light guides and PMTs.

%%%%%%%%%%%%%%%%%%%%%%%%%%%%%%%%%%%%%%%%%%%%%%%%%%%%%%%%%%%%%%%%%%%%%%%%%%%%%%%%%%%%%%%%%%%
% Absorber
%%%%%%%%%%%%%%%%%%%%%%%%%%%%%%%%%%%%%%%%%%%%%%%%%%%%%%%%%%%%%%%%%%%%%%%%%%%%%%%%%%%%%%%%%%%
\subsection{Absorber}

The absorber plate is required to be made of appropriate atoms to meet the following aspects.
The lifetime of muonic atoms in the absorber, \ltime{short}, is required to be significantly shorter than that of BG DIO, $\sim 2~\usec$.
On the other hand, it must be longer than 100~\nsec{} to be well separated from the beam-induced prompt BG hits in the \T1 and \T2 Counters.

Copper and aluminium were considered as candidate materials.  
The absorber lifetime, \ltime{short}, and the decay-in-orbit (DIO) rate, \R{DIO}, were averaged from experimental records summarised in~\cite{MuonicAtomLife.1}, resulting in $164\pm3~\nsec$ and $7.20\pm0.04\%$ for copper, and $878\pm19~\nsec$ and $39.07\pm0.04\%$ for aluminium~\cite{MuonicAtomLife.1,MuonicAtomLife.2,MuonicAtomLife.3,MuonicAtomLife.4,MuonicAtomLife.5,MuonicAtomLife.6}.  
A prototype RC equipped with copper and aluminium absorbers was developed and tested in June 2022 at the D2 muon beamline of the Materials and Life Science Experimental Facility (MLF) at J-PARC~\cite{MLF}.  
The thicknesses were 0.5~mm for copper and 1.6~mm for aluminium, adjusted to achieve the same weight per unit area, which corresponds to the lowest target momentum range of 30--50~\MeVc{} when used with the \T0 Counter and no degrader.
The MLF provides a double-pulsed negative muon beam produced from a carbon target by the J-PARC 3~GeV proton beam, with characteristics essentially similar to the slow muon beam to be measured in \PhaseAlpha{}, including beam BGs.

\begin{figure}[tb]
    \centering
    \begin{minipage}[b]{0.45\linewidth}
        \centering
        \includegraphics[width=\textwidth]{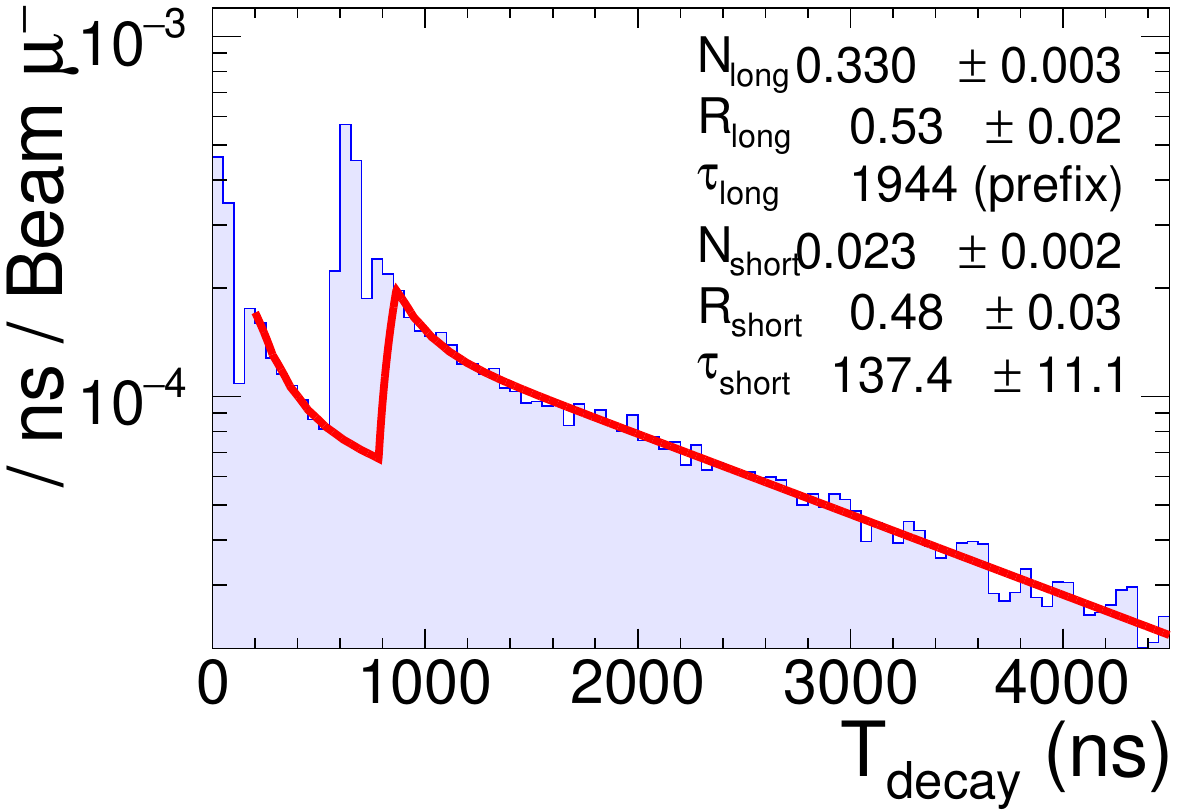}
        \subcaption{Copper}
    \end{minipage}
    \begin{minipage}[b]{0.45\linewidth}
        \centering
        \includegraphics[width=\textwidth]{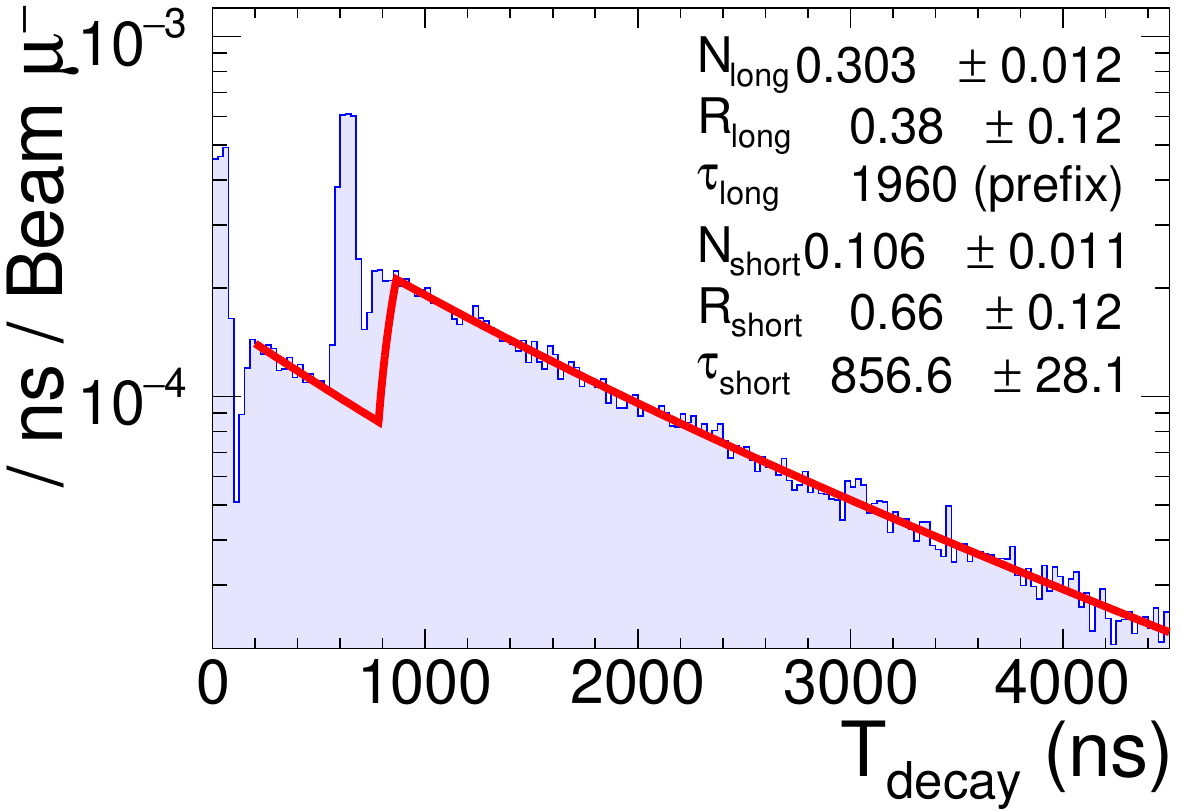}
        \subcaption{Aluminium}
    \end{minipage}
    \caption{
        Comparison of copper and aluminium as absorber materials in the experiment at the Material and Life Science Experimental Facility in J-PARC.
        The vertical axis represents the number of entries per nanosecond, normalised to the number of injected muons.
        The peaks around 0 and 600~\nsec{} correspond to background hits from double bunches of injected beam particles.  
        The red line shows the fit result of \eq{eq:MLFFitFunc}, with the parameters indicated in the legend. 
        The regions around the peaks were excluded from the fit.   
    }
    \label{fig:CuAlComparison}
\end{figure}

\fig{fig:CuAlComparison} shows the reconstructed $T_{\rm decay}$ distributions for both absorbers.  
After each of the two peaks from muon injection, decay curves are superimposed.  
The distributions are fitted with the following function derived from \eq{eq:Tdecay},
\begin{equation}
    \begin{split} 
        F'(T_{\rm decay}) &= \sum_{i={\rm short,long}} 
        \frac{\N{i} R_i}{\ltime{i}} \exp\left(-\frac{T_{\rm decay}}{\ltime{i}}\right) \\
        &\qquad + u(T'_{\rm decay})
        \frac{\N{i}(1 - R_i)}{\ltime{i}} \exp\left(-\frac{T'_{\rm decay}}{\ltime{i}}\right)
    \end{split},
\label{eq:MLFFitFunc}
\end{equation}
where $T'_{\rm decay} = T_{\rm decay} - 800~\nsec$, and $u(T_{\rm decay})$ is a step function.
The regions around the peaks are excluded from the fit.
The fit results are also indicated in the figure legend, whereas the long lifetime, \ltime{long}, is fixed from a separate pre-fit to the tail part.

DIO in copper exhibits a more distinct component in the distribution than in aluminium, although it was initially expected to have a low signal-to-noise ratio due to its lower \R{DIO}.
Good separation requires including a sufficiently long time in the fitting, such as the drawn region after the second peak.  
However, the interval of the beam bunches for COMET is 1.17~\usec{}, which is insufficient for such an approach. 
For copper, the short component separation was found to be easier, even with the 600~\nsec{} interval of the D2 beamline.  
In conclusion, a copper plate of $300\times 300\times 0.5~\mmp3$ was chosen as the absorber.

%%%%%%%%%%%%%%%%%%%%%%%%%%%%%%%%%%%%%%%%%%%%%%%%%%%%%%%%%%%%%%%%%%%%%%%%%%%%%%%%%%%%%%%%%%%
% T1 and T2 Counters
%%%%%%%%%%%%%%%%%%%%%%%%%%%%%%%%%%%%%%%%%%%%%%%%%%%%%%%%%%%%%%%%%%%%%%%%%%%%%%%%%%%%%%%%%%%
\subsection{$T_1$ and $T_2$ Counters}

Both of the \T1 and \T2 Counters need to effectively detect penetrating DIO electrons from the absorber plate.
Their dimensions need to follow those of the \Tz2 Counter; they should not be too large to minimise hits by the off-centre beam-induced BG particles.
The \T2 Counter can be a little wider than the \T1 Counter so that the solid-angle coverage increases.

In contrast to the \T0 counter, they can be thick enough to collect a sufficient amount of scintillation light with a single PMT because they do not need to transmit the low-momentum muons.
Additionally, the thickness of the \T1 Counter should be balanced between preventing BG protons from reaching the \T2 Counter and accepting as many DIO electrons emitted with a large angle as possible.
We chose $200\times 200\times 10~\mmp{3}$ and  $300\times 300\times 5~\mmp{3}$ for the \T1 and \T2 Counters, respectively.

A 1-cm thick plastic scintillator stops protons with kinetic energies up to $\sim 35$~MeV~\cite{NISTPSTAR}.
Although experimental data on the kinetic energy of charged particles produced by \mucap{} in copper are limited, the fraction exceeding 40~MeV has been reported to be 0.2\% of all \mucap{} events~\cite{MuonCaptureSecondaries}, which is sufficiently small compared to $\R{DIO} = 7.2\%$.

%%%%%%%%%%%%%%%%%%%%%%%%%%%%%%%%%%%%%%%%%%%%%%%%%%%%%%%%%%%%%%%%%%%%%%%%%%%%%%%%%%%%%%%%%%%
% Assembly
%%%%%%%%%%%%%%%%%%%%%%%%%%%%%%%%%%%%%%%%%%%%%%%%%%%%%%%%%%%%%%%%%%%%%%%%%%%%%%%%%%%%%%%%%%%
\subsection{Assembly}

\begin{figure}[tb]
    \centering
    \begin{minipage}[b]{\columnwidth}
        \centering
        \includegraphics[height=0.3\textheight]{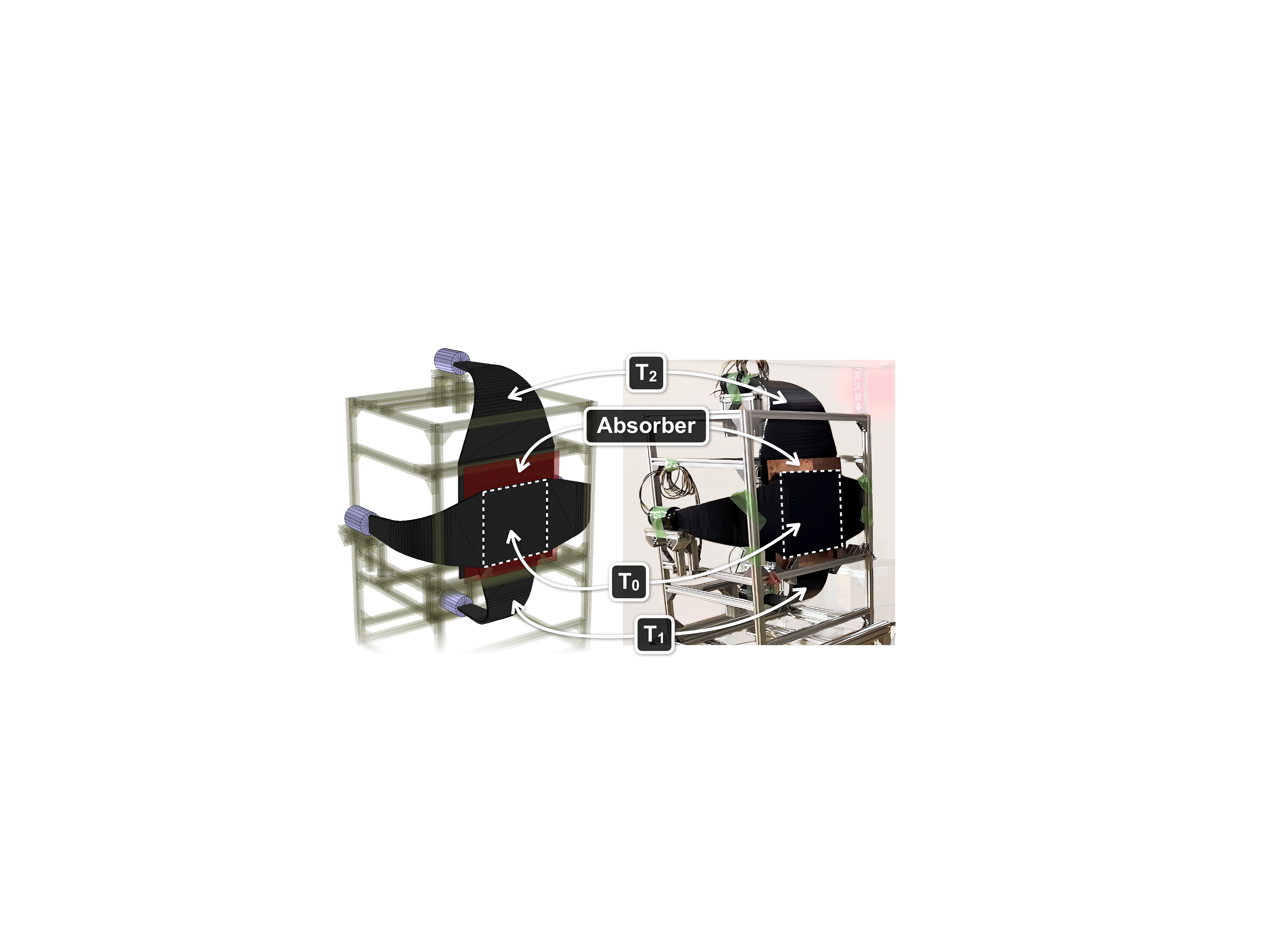}
        \subcaption{(Left) Three-dimensional model and (right) photograph of the Range Counter assembly}
    \end{minipage}
    \vspace{1mm}
    
    \begin{minipage}[b]{\columnwidth}
        \centering
        \includegraphics[height=0.2\textheight]{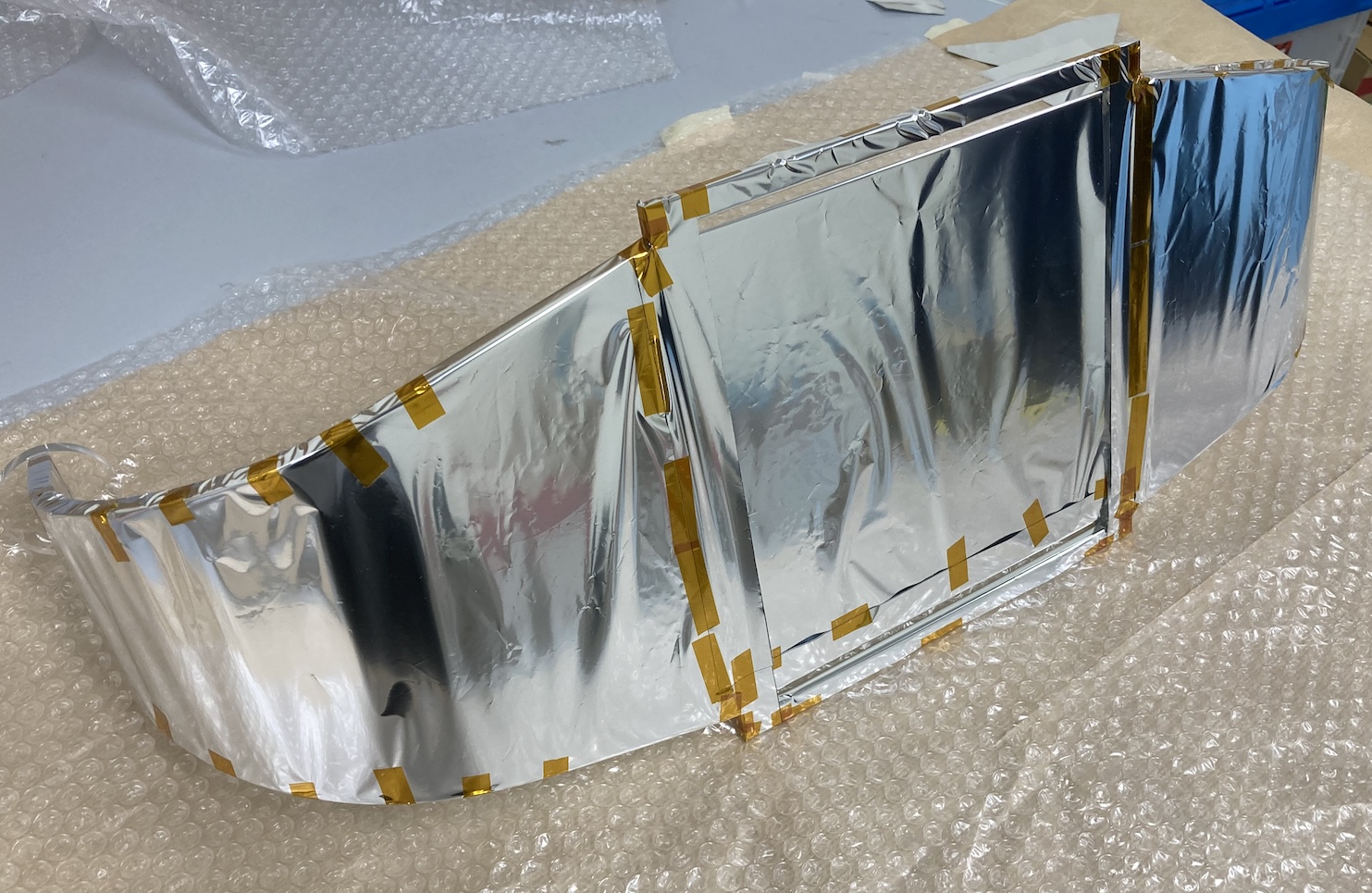}
        \subcaption{\T0 Counter without light shield}
    \end{minipage}
    \caption{
        Constructed Range Counter assembly with a three-dimensional model and the 20~cm-square \T0 Counter before being wrapped for light shielding. 
    }
    \label{fig:RCAssembly}
\end{figure}

We constructed all the timing counters and assembled them based on the described design.
\fig{fig:RCAssembly} shows photos of the assembled RC, along with a 3D illustration for clarity, and the \Tz2 Counter before being wrapped in a black sheet.
The counters are tightly aligned with the copper absorber.
To prevent interference among the counters, the light guides extend laterally for the \T0 Counter and vertically for the \T1 and \T2 Counters.
Supporting beams were installed as needed to avoid placing mechanical stress on the PMT contact.

%%%%%%%%%%%%%%%%%%%%%%%%%%%%%%%%%%%%%%%%%%%%%%%%%%%%%%%%%%%%%%%%%%%%%%%%%%%%%%%%%%%%%%%%%%%
% Evaluation
%%%%%%%%%%%%%%%%%%%%%%%%%%%%%%%%%%%%%%%%%%%%%%%%%%%%%%%%%%%%%%%%%%%%%%%%%%%%%%%%%%%%%%%%%%%
\subsection{Estimation of the purity and acceptance}

% Explanation of the parameters
The key performance parameters determined by the geometrical setup, namely \pur{DIO} and \accep{DIO} in \eq{eq:DIO}, were estimated using a Monte Carlo simulation with Geant4 version 10.6.p3 and the QGSP\_BERT physics list~\cite{Geant4a, Geant4b, Geant4c, Geant4d}.

% Simulation setup
The absorber plate and all the timing counters including their reflector and light-shield wrappings are constructed and placed in the simulation setup as designed.
The light guides and PMTs are not included as their geometrical influences to this evaluation are negligible.
The simulation generates muons with random momentum directed to the RC, ensuring uniform stopping distribution within the absorber.
The beam-spreading size is larger than the RC's lateral size, as expected in \PhaseAlpha{}.
The stopped muons decay and emit secondary particles, and the resulting true energy deposits in the \T1 and \T2 Counters are recorded with no fluctuation.
The analysis selects the events in which the muon hits the \T0 Counter and stops in the absorber, and collects simultaneous hits originating from DIO electrons or \mucap-induced particles.

\begin{figure}[tb]
    \centering
    \begin{minipage}{0.32\textwidth}
        \centering
        \includegraphics[width=\textwidth]{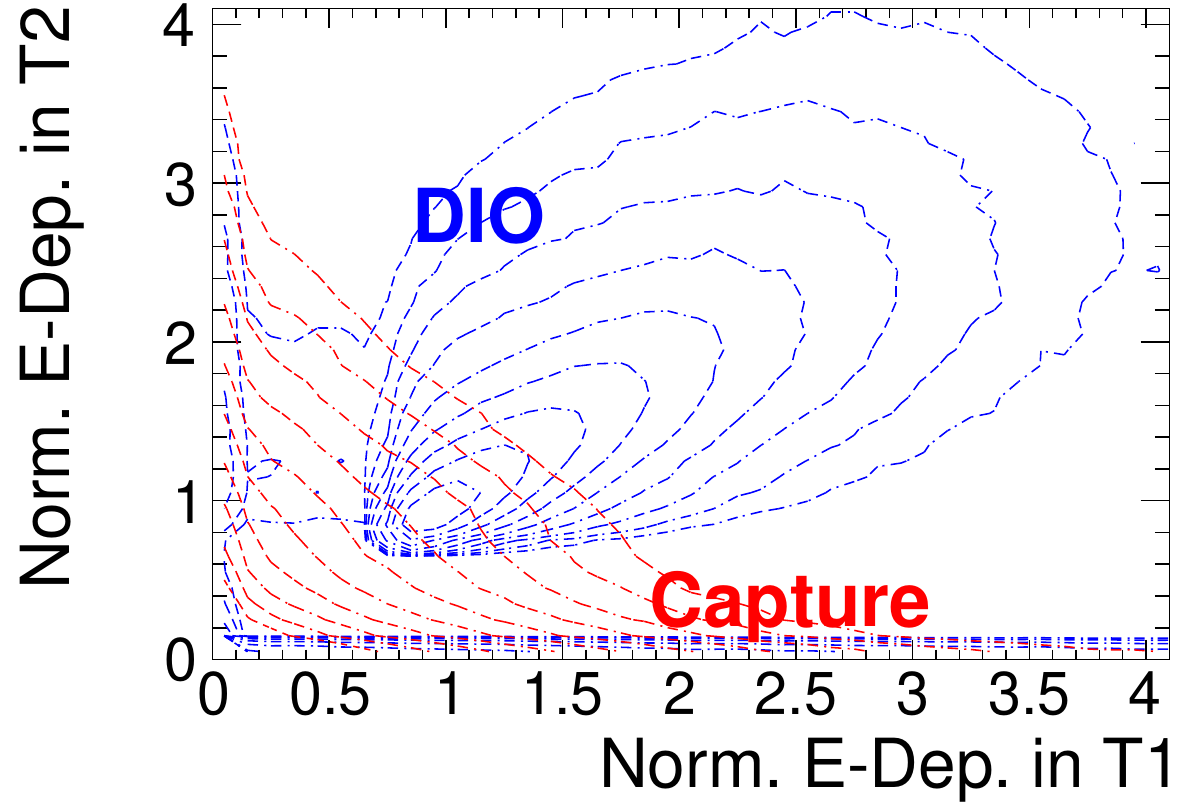}
        \subcaption{Energy deposit: contour plots}
    \end{minipage}
    \hfill
    \begin{minipage}{0.32\textwidth}
        \centering
        \includegraphics[width=\textwidth]{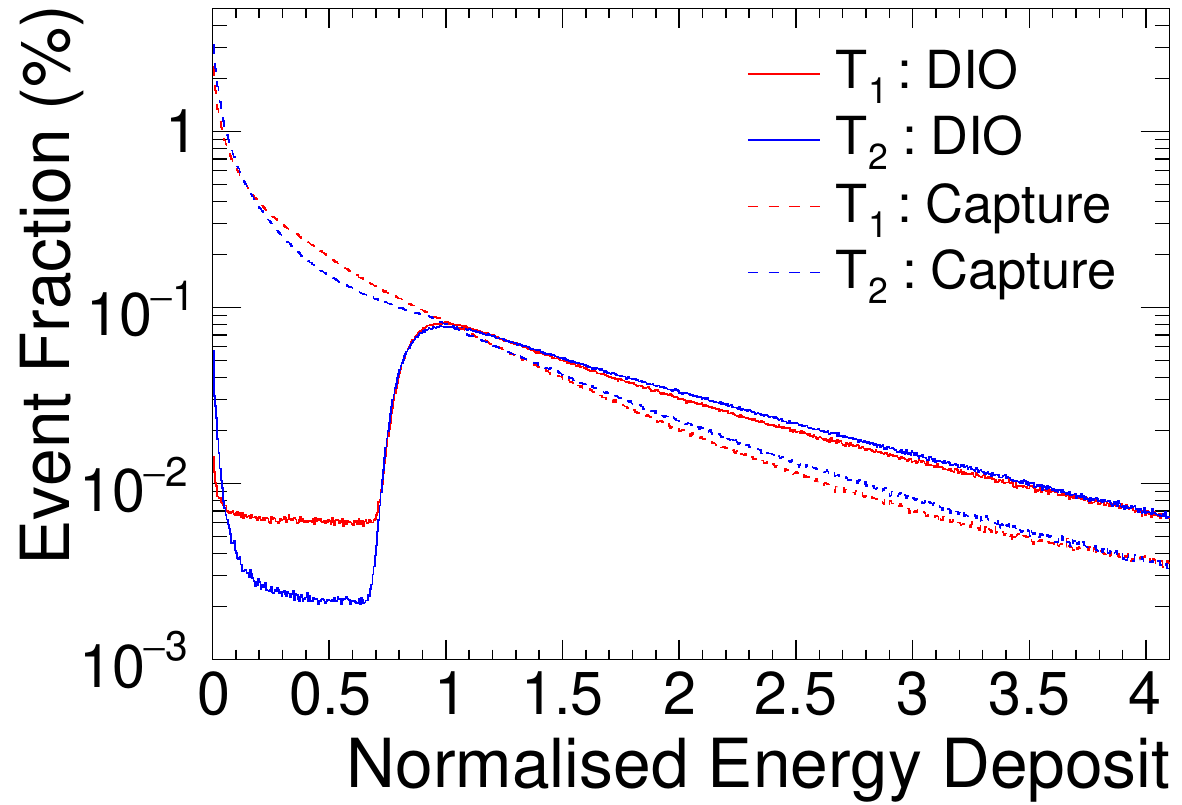}
        \subcaption{Energy deposit: projected plots}
    \end{minipage}
    \hfill
    \begin{minipage}{0.32\textwidth}
        \centering
        \includegraphics[width=\textwidth]{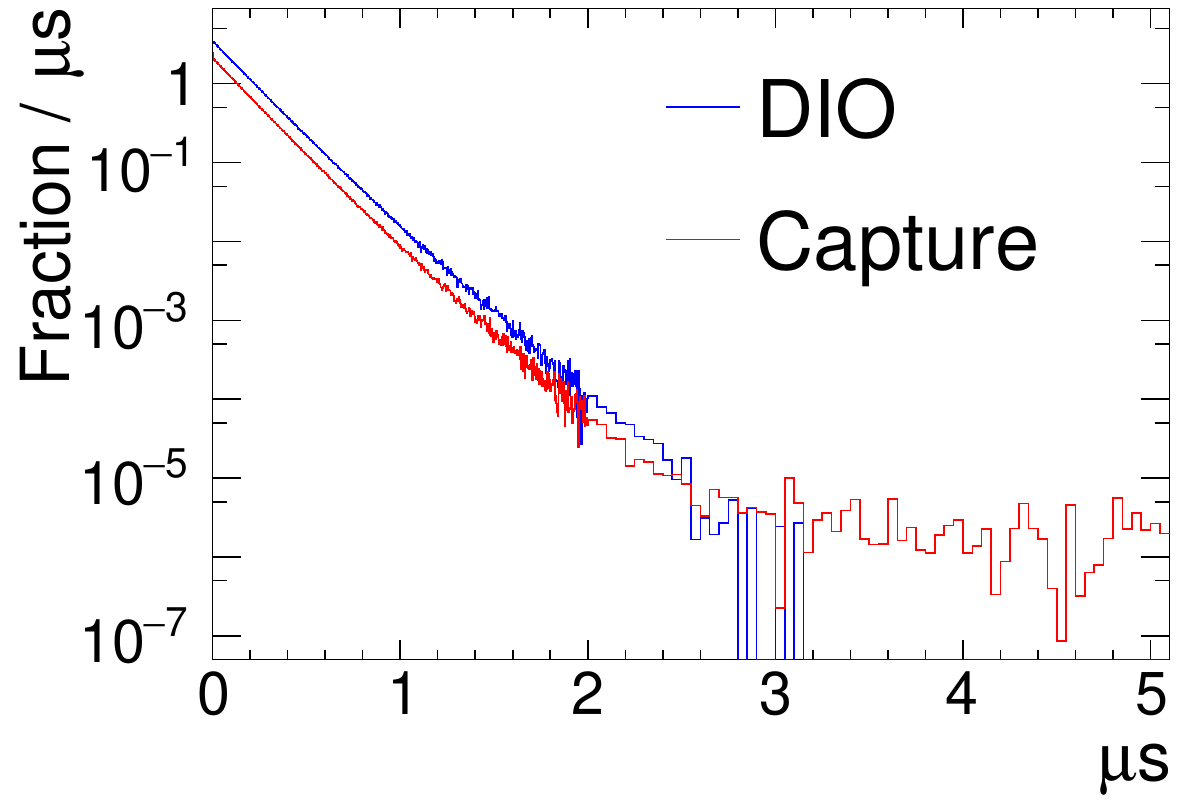}
        \subcaption{Hit time}
    \end{minipage}
    \caption{
        (a,b) Energy deposits in the \T1 and \T2 Counters by DIO electrons and \mucap-induced particles that hit the counters simultaneously and (c) their timing distributions, simulated by Geant4.
        The energy deposits are normalised to the most probable value by DIO electrons for each counter.
        The time bin widths of (c) become wider after 2~\usec{} due to limited statistics.
    }
    \label{fig:DIOCap}
\end{figure}

\fig{fig:DIOCap}(a,b) displays the energy deposit distributions in the counters for DIO electrons and \mucap-induced background particles that simultaneously hit the counters.
The energy deposits are normalised to the most probable value (MPV) of the DIO electron deposits in each counter.
The two contour distributions show different correlations.
The coincidental energy deposits associated with \mucap{} are mainly caused by $e^\pm$ and $\gamma$-rays, which originate from the interactions of $\gamma$-rays and neutrons.
Their momenta are typically too low to produce the uncorrelated energy deposits characteristic of DIO electrons.
This justifies the application of energy cuts to isolate signal events.
In \fig{fig:DIOCap}(a), DIO electrons still show entries below 0.7 in both counters due to geometrical effects.

The time distributions of the coincidental hits are shown in \fig{fig:DIOCap}(c).
Although both decay curves appear consistent with the same lifetime as muonic atoms in copper, \mucap-induced hits has a slightly longer effective lifetime, as secondary slow neutrons can persist over longer timescales.
There is also a small fraction of events with an even longer lifetime, appearing after 3~\usec{}, but this is negligible for the evaluation of \pur{DIO}.

The slight difference between these lifetimes cannot be resolved when reconstructing \N{short} using \eq{eq:Tdecay}.
To evaluate \pur{DIO} in terms of the components that genuinely contribute to \N{short}, $\N{DIO}^{\rm cut}$ and $\N{cap}^{\rm cut}$ are extracted by fitting an exponential function to each distribution after applying the cut, within the time range from 0.1 to 1.0~\usec.

\begin{figure}[tb]
    \centering
    \begin{minipage}{0.48\columnwidth}
        \centering
        \includegraphics[width=\textwidth]{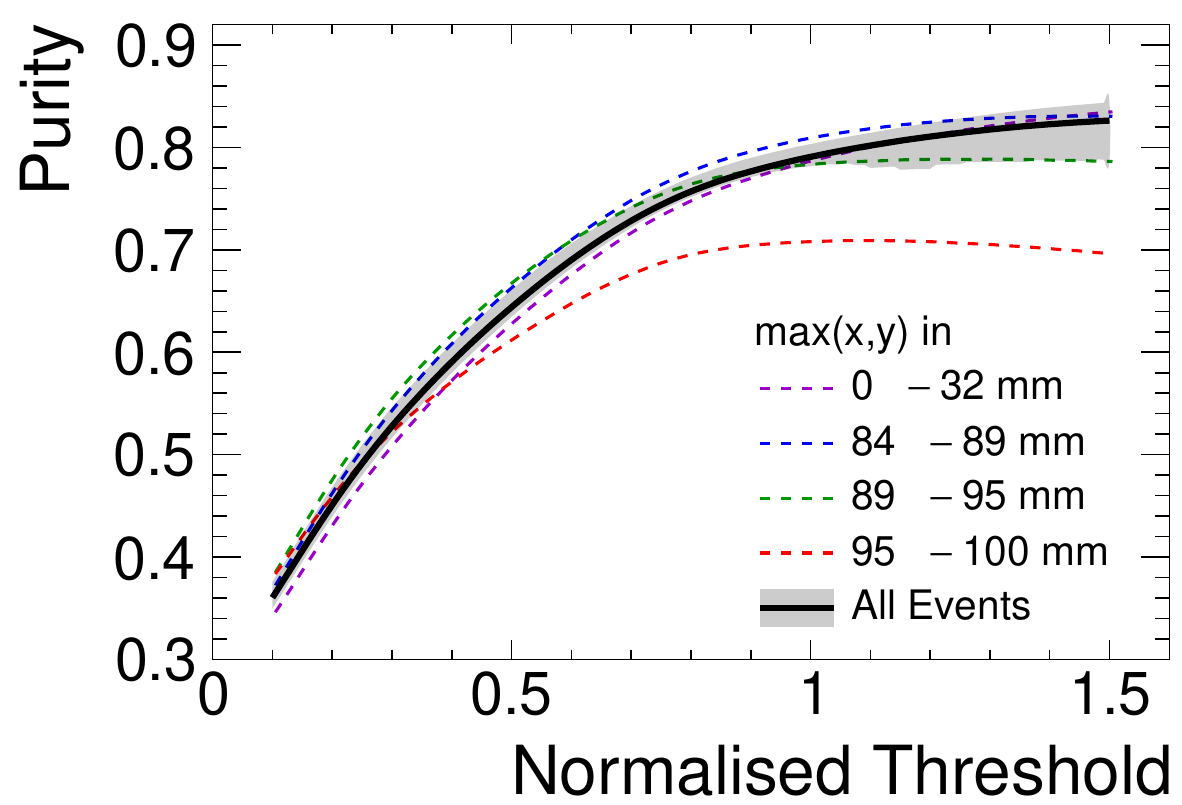}
        \subcaption{\pur{DIO}: Purity}
    \end{minipage}
    \hfill
    \begin{minipage}{0.48\columnwidth}
        \centering
        \includegraphics[width=\textwidth]{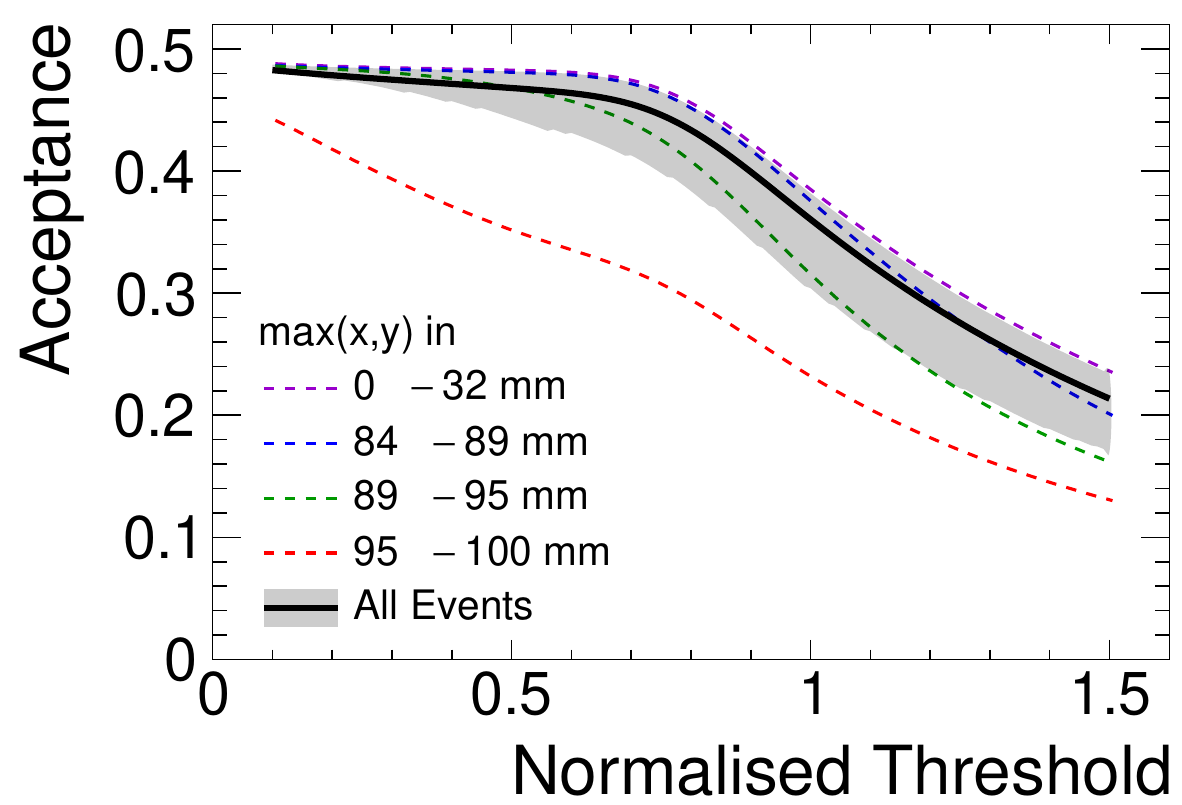}
        \subcaption{\accep{DIO}: Acceptance}
    \end{minipage}
    \caption{
        Purity and acceptance of the \T1 and \T2 Counters for DIO electrons as functions of the energy cut threshold, evaluated by simulation.
        The threshold is normalised to the most probable energy deposit by DIO electrons.
        Dashed lines correspond to different ranges of the muon-stopping position, characterised by $\max(x, y)$, where $x$ and $y$ are the horizontal and vertical distances from the absorber centre.
        Solid lines represent the results over the entire absorber area, with grey bands indicating variations due to the stopping position.
    }
    \label{fig:DIOCap_Result}
\end{figure}

\fig{fig:DIOCap_Result} shows the evaluated \pur{DIO} and \accep{DIO} as functions of the normalised threshold of the energy cut.
Since the energy deposit distributions in \fig{fig:DIOCap}(a,b) for both counters exhibit similar shapes along the normalised energy axis, a unified threshold was applied to both counters.
The solid lines represent the results using all events, whereas the dashed lines correspond to different ranges of the muon-stopping position, denoted by $D = \max(x,y)$, where $x$ and $y$ are the horizontal and vertical distances from the centre of the absorber.
Both curves drop steeply beyond $D \sim 95~\mm$ due to the geometrical limitations of the setup.
Systematic uncertainties are evaluated from the root-mean-square fluctuations around the solid lines, considering the dependence on the stopping position up to $D = 95~\mm$.

\accep{DIO} remains stable at around 47\% up to a threshold of 0.4.
In contrast, \pur{DIO} increases more sharply, indicating that a higher threshold is preferable.
Still, the optimum threshold should be determined based on the actual performance of the \T1 and \T2 Counters, namely \eff{DIO}.
It must not exceed the point where \eff{DIO} becomes unstable due to the finite energy resolutions of the counters.
As discussed later in \sect{sec:Performance}, a threshold of 0.4 is conservatively achievable, at which \pur{DIO} and \accep{DIO} are estimated to be $59.6^{+1.8}_{-1.1}\%$ and $47.1^{+1.2}_{-1.5}\%$, respectively.

%%%%%%%%%%%%%%%%%%%%%%%%%%%%%%%%%%%%%%%%%%%%%%%%%%%%%%%%%%%%%
% Performance
%%%%%%%%%%%%%%%%%%%%%%%%%%%%%%%%%%%%%%%%%%%%%%%%%%%%%%%%%%%%%
\section{Performance}
\label{sec:Performance}

The performance evaluation focuses on \eff{trig} and \eff{DIO}.
To this end, an experiment was conducted in November 2023 at the \piMone{} beamline of Paul Scherrer Institut in Switzerland.
The beamline provides a beam of $\pi^-$, $\mu^-$, and $e^-$ with momenta of 100--300~\MeVc.

\begin{figure}[tb]
    \centering
    \includegraphics[width=\columnwidth]{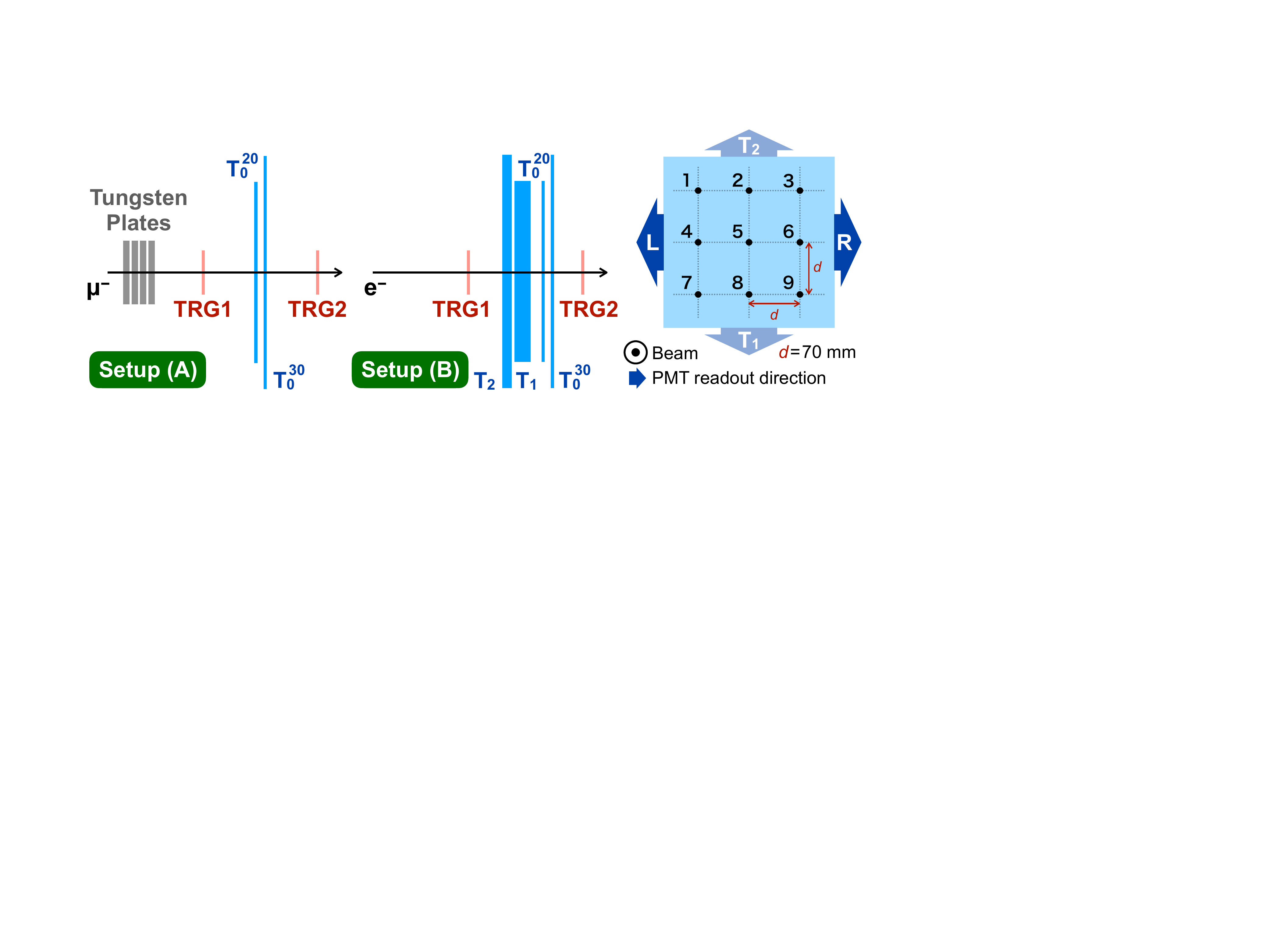}
    \caption{
        Experimental setups at the \piMone{} beamline of Paul Scherrer Institut.
        Trigger counters (TRG) detect the beam and trigger data acquisition.
        Tungsten plates are used to control the momentum of the beam muons entering the tested counters.
        The right figure shows the nine predefined beam spots placed at 70~mm intervals, with point 5 at the centre.
        The lateral arrows labelled R(ight) and L(eft) denote the readout orientations of the PMTs of the \T0 Counters with respect to the beam axis, while each vertical arrow indicates that for the \T1 and \T2 Counters.
    }
    \label{fig:PSISetup}
\end{figure}

% Experimental setting
The experimental setups are illustrated in \fig{fig:PSISetup}, where Setups (A) and (B) correspond to the measurements of \eff{trig} and \eff{DIO}, respectively.
The \Tz2, \Tz3, \T1, and \T2 Counters were positioned between a pair of trigger counters, TRG1 and TRG2, made of thin plastic scintillator with dimensions of $20 \times 20 \times 0.5~\mmp3$, which were used to initiate DAQ.
Each trigger counter was wrapped in the same material as the main timing counters and optically coupled, via a 20~mm-wide light guide bar, to a PMT using the same optical grease.
The PMTs for all counters were operated at the nominal voltages recommended by Hamamatsu.
The \T0 Counters in Setup~(B) were installed with the expectation that they might serve as future reference data, although their data fall outside the scope of this study.

The assembly was mounted on a three-dimensional movable stage.  
Measurements were performed at nine beam spots arranged in a 7~cm-pitch grid on the counters.  
The TRG counters effectively defined the beam spot size as the beam was not purposely focused.  
The \T0 Counters are equipped with dual PMTs on both the left and right sides, and their readout orientations with respect to the beam axis are illustrated in \fig{fig:PSISetup}.
In contrast, each of the \T1 and \T2 Counters has a single PMT on the bottom and top sides, respectively.

% Beam setting
The beam momentum was fixed at 250~\MeVc{} in both setups.
In Setup (A), tungsten plates were used as muon momentum degraders to adjust the momentum of muons incident on the counters.
\tab{tab:Momenta} summarises all the configurations with the estimated momentum ranges.
Due to limited beam time, some configurations were only tested at the central beam spot.

The listed ranges are estimated by Geant4-based simulation in which 250~\MeVc{} negative muons are injected into tungsten plates of various total thicknesses, and their momenta after passing through the degrader are sampled at the tested counters.
The mean values vary from 247.3 to 59.0~\MeVc{} depending on the degrader thickness up to 56~mm, while the momentum spread stays limited to $\sim 10~\MeVc$.

\begin{table}[tb]
    \centering
    \caption{
        Total thickness of tungsten degrader plates in Setup (A) and the resulting momentum and fluctuation, estimated by Geant4-based simulation.
        Mean values correspond to the most probable values (MPVs), and uncertainties represent the root mean square of deviations above and below each MPV.
        The beam spots irradiated in each setup, defined in \fig{fig:PSISetup}, are also listed.
    }
    \label{tab:Momenta}
    \begingroup
        \renewcommand{\arraystretch}{1.3}
        \footnotesize
        \begin{minipage}{0.4\columnwidth}   
            \centering
            \begin{tabular}{rll}
                \toprule
                \noalign{\vspace{-3pt}}
                mm & \multicolumn{1}{c}{\MeVc} & Spot\\
                \noalign{\vspace{-2pt}}
                \midrule
                0  & $247.3^{+0.1}_{-0.6}$ & 1--9\\
                30 & $169.2^{+2.4}_{-3.8}$ & 5\\
                40 & $136.4^{+3.6}_{-4.9}$ & 5\\
                48 & $106.8^{+4.3}_{-7.9}$ & 1--9\\
                \bottomrule
            \end{tabular}
        \end{minipage}
        \begin{minipage}{0.4\columnwidth}   
            \centering
            \begin{tabular}{rll}
                \toprule
                \noalign{\vspace{-3pt}}
                mm & \multicolumn{1}{c}{\MeVc} & Spot\\
                \noalign{\vspace{-2pt}}
                \midrule
                50 & $96.4^{+5.2}_{-9.4}$ & 1--9\\
                52 & $85.0^{+6.2}_{-12}$ & 5\\
                54 & $73.0^{+7.1}_{-14}$ & 1--9\\
                56 & $59.0^{+8.2}_{-12}$ & 5\\
                \bottomrule
            \end{tabular}
        \end{minipage}
    \endgroup
\end{table}

\begin{figure}[tb]
    \centering
    \includegraphics[width=\columnwidth]{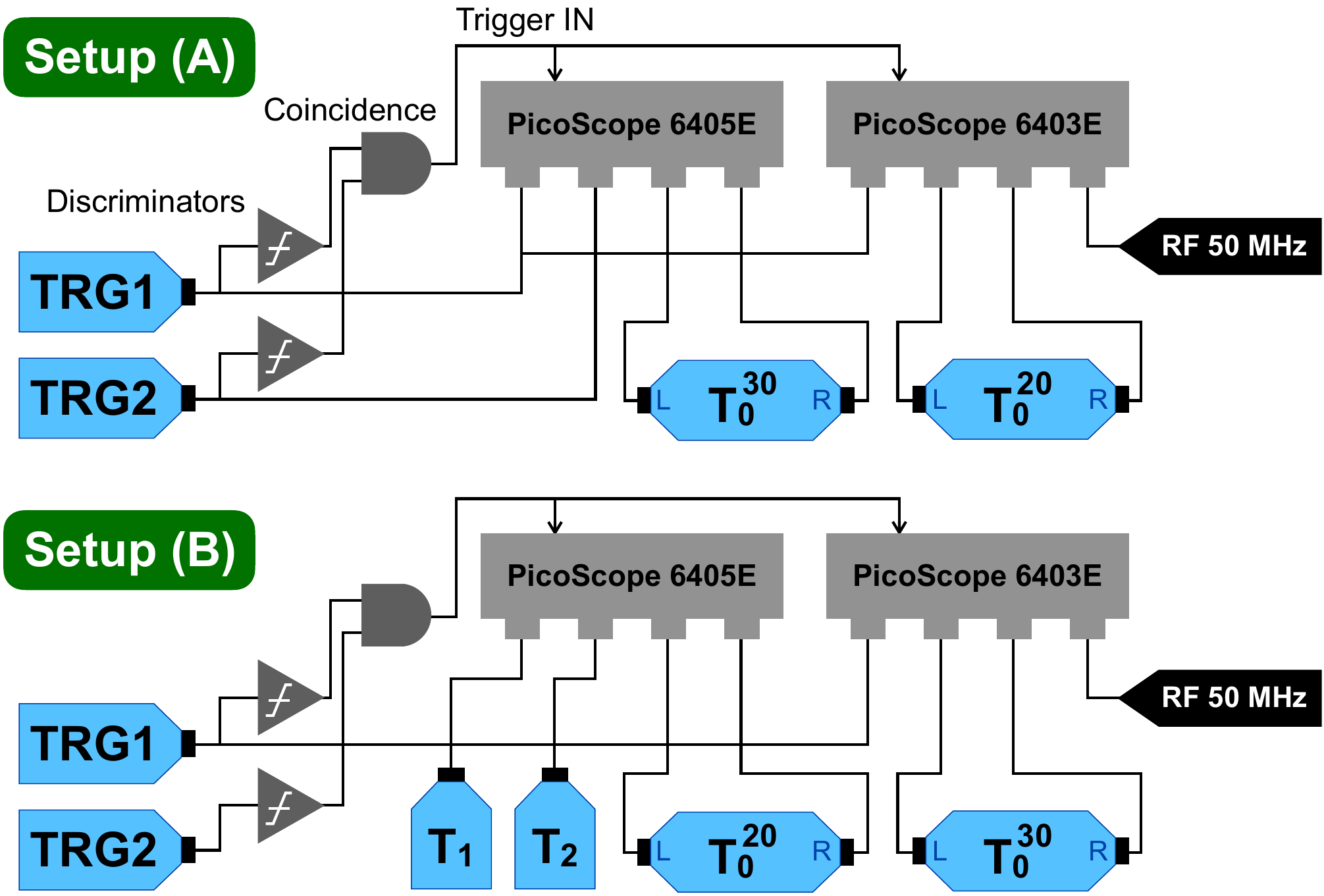}
    \caption{
        Data acquisition setups at the \piMone{} beamline of the Paul Scherrer Institut, corresponding to \fig{fig:PSISetup}.
        Waveforms from the counters were recorded using PicoScope 6000E series digitisers.
        A coincidence signal from the two trigger counters (TRG) was used to trigger the digitisation.
        A 50~MHz sine-wave signal driving the accelerator was also recorded to enable time-of-flight measurements for beam particle identification.
    }
    \label{fig:PSIDAQ}
\end{figure}

% DAQ
\fig{fig:PSIDAQ} shows the DAQ schematics for both setups.
Conventional NIM-standard modules discriminated hits in the TRG counters and generated a trigger signal when simultaneous hits were detected.
Waveforms from the tested counters were recorded for waveform analysis using digitiser modules, Pico Technology PicoScope 6405E and 6403E; their bandwidths are 500 and 300~MHz, respectively, and the digitising resolution is 8 bits.
The used sampling period was 0.8~\nsec, and the dynamic ranges were adjusted to to accommodate most of the signal waveform amplitudes.

A 50~MHz sine-wave signal driving the accelerator was also recorded to enable particle identification.
Its phase was synchronised with the beam injection timing, so that the timing difference between this signal and the TRG counter hits corresponds to the particles’ time of flight (TOF) from the primary target along the 21~m beamline.
Beam muons and electrons were identified based on their TOF and the energy deposit measured by the TRG1 counter.

%%%%%%%%%%%%%%%%%%%%%%%%%%%%%%%%%%%%%%%%%%%%%%%%%%%%%%%%%%%%%%%%%%%%%%%%%%%%%%%%%%%%%%%%%%%%
% Muon-trigger Efficiency
%%%%%%%%%%%%%%%%%%%%%%%%%%%%%%%%%%%%%%%%%%%%%%%%%%%%%%%%%%%%%%%%%%%%%%%%%%%%%%%%%%%%%%%%%%%%
\subsection{Muon-trigger Efficiency}

\eff{trig} represents the probability that the \T0 Counter detects a passing muon.
In \PhaseAlpha{}, trigger signals are generated by discriminating signal waveforms from the \T0 Counter with a fixed threshold.
In accordance with that, the same procedure and logic are applied to the offline waveform analysis in this evaluation.
Thus, \eff{trig} is evaluated as a function of the muon momentum and the discrimination threshold.
For that purpose and better compatibility, the pulse amplitude and threshold are normalised to the MPV of the distribution of pulse peak amplitudes obtained for each \T0 Counter when the beam muons hit the central beam spot with no degrader. 

\begin{figure}[tb]
    \centering
    \begin{minipage}{0.48\columnwidth}
        \centering
        \includegraphics[width=\textwidth]{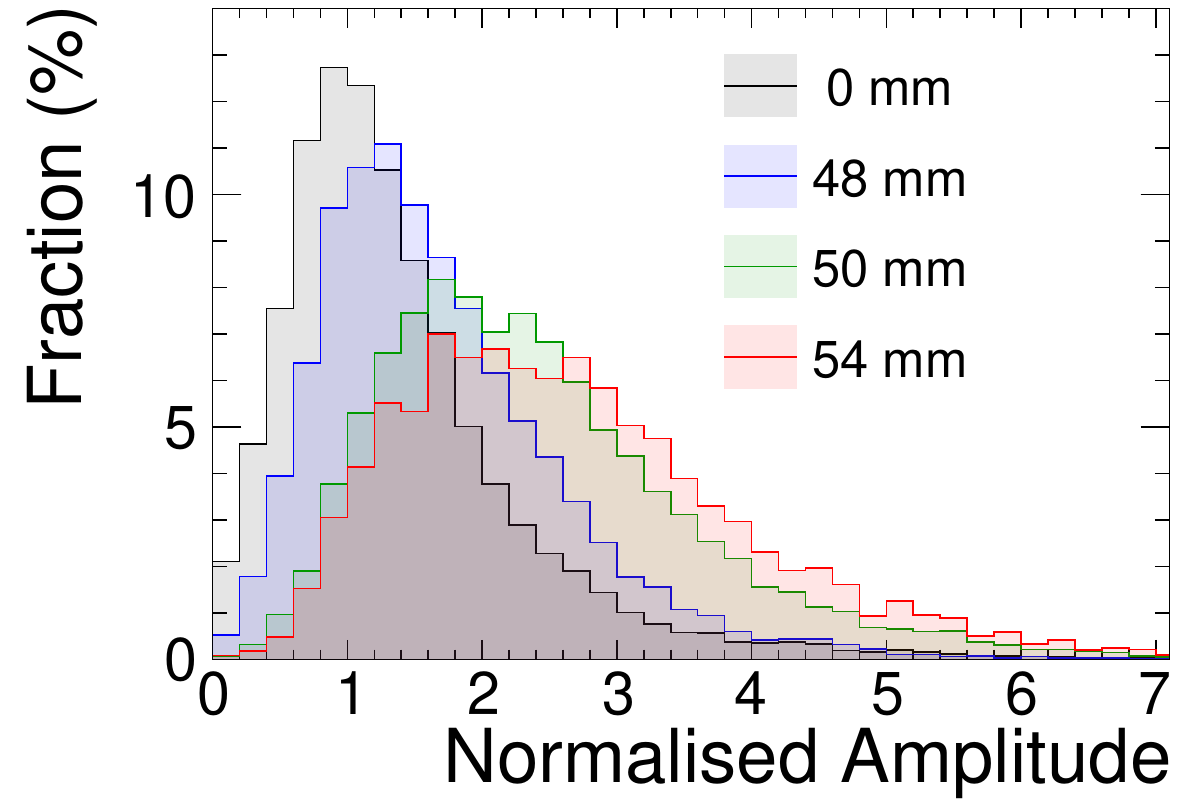}
        \subcaption{\Tz2 Counter}
    \end{minipage}
    \hfill    
    \begin{minipage}{0.48\columnwidth}
        \centering
        \includegraphics[width=\textwidth]{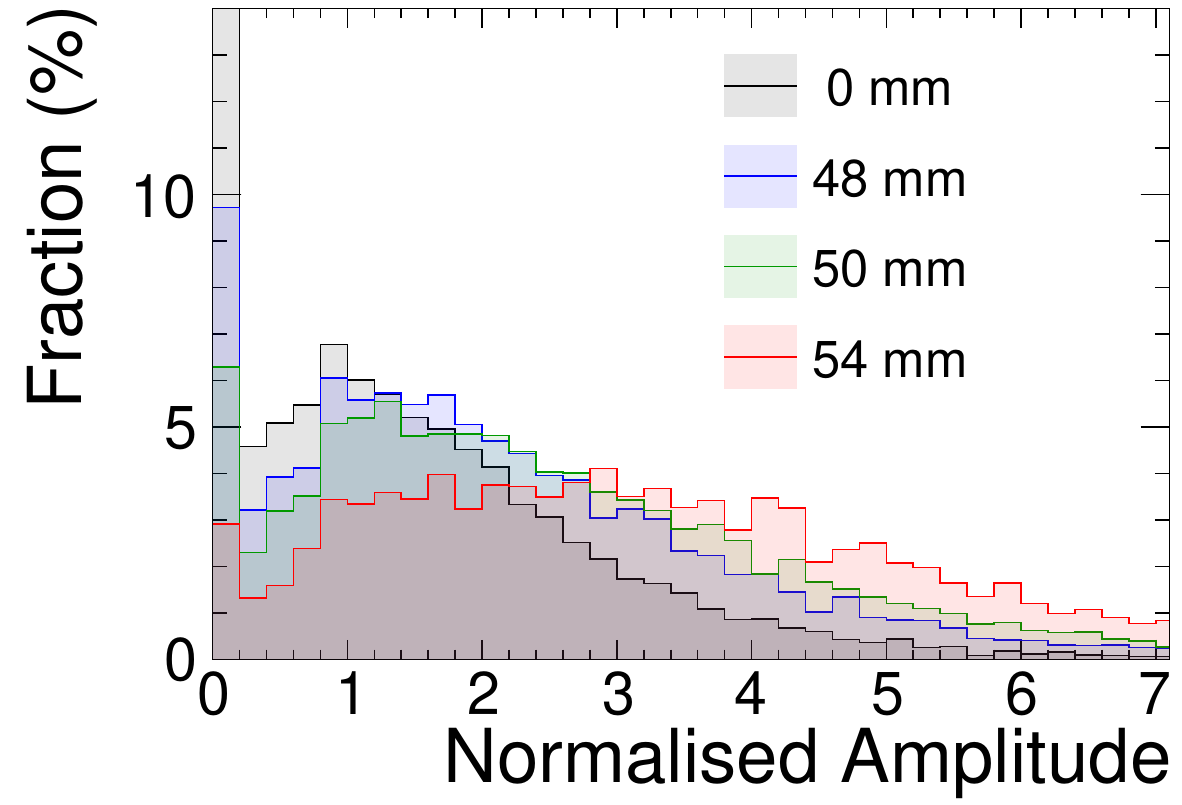}
        \subcaption{\Tz3 Counter}
    \end{minipage}
    \caption{
        Distributions of the pulse peak amplitudes at the central beam spot obtained from the right readout side for the \T0 Counters for different degrader thicknesses.
        Waveform amplitudes are normalised to the most probable pulse peak amplitudes for each counter when no degrader (0 mm) is used.
        The vertical axes are normalised to the total entries.
    }
    \label{fig:WaveformAmplitude}
\end{figure}

% Waveform amplitude
\fig{fig:WaveformAmplitude} shows the distributions of pulse peak amplitudes at the central beam spot for the \T0 Counters under different degrader configurations, as measured from the right-hand readout side.  
The aforementioned normalisation is applied to the amplitudes, and the distributions with no degrader (0~mm) exhibit their MPVs at one.  
It is evident that the \Tz3 Counter is no longer capable of efficiently detecting single photoelectrons, as indicated by the large number of entries in the leftmost bin.  
This was the critical reason for abandoning the 30~cm-square scintillator as a candidate for the \T0 Counter.  
We have confirmed, through a test using collimated $\beta$-rays from a $\rm{}^{90}Sr$ source in the laboratory, that these reference MPVs correspond to approximately 3.8 and 1.5 photoelectrons for the \Tz2 and \Tz3 Counters, respectively.  

The pedestal noise width was independently evaluated from a noise region in the waveform data and found to be approximately 0.1~p.e.\ ($1\sigma$ of a Gaussian), corresponding to 0.04 and 0.1 on the normalised scale for the \Tz2 and \Tz3 Counters, respectively.  
Accordingly, we adopt a minimum threshold of 0.2 to maintain a reasonable signal-to-noise ratio for the \Tz2 Counter.  
This value is somewhat aggressive for the \Tz3 Counter, but we continue to apply the same threshold to ensure methodological consistency between the two counters.  

Still, random 1~p.e.\ noise pulses from the PMT can occasionally exceed this threshold, which corresponds to approximately 0.8~p.e.  
Nevertheless, such noise pulses do not contribute to the background in the short component of the $T_{\rm decay}$ distribution.  
From a practical point of view, the use of timing coincidences within a narrow time window, defined by signals from other detectors or by the accelerator's injection timing, can further reduce the probability of noise-induced triggers.

\begin{figure}[tb]
    \centering
    \begin{minipage}{0.47\textwidth}
        \centering
        \includegraphics[width=\textwidth]{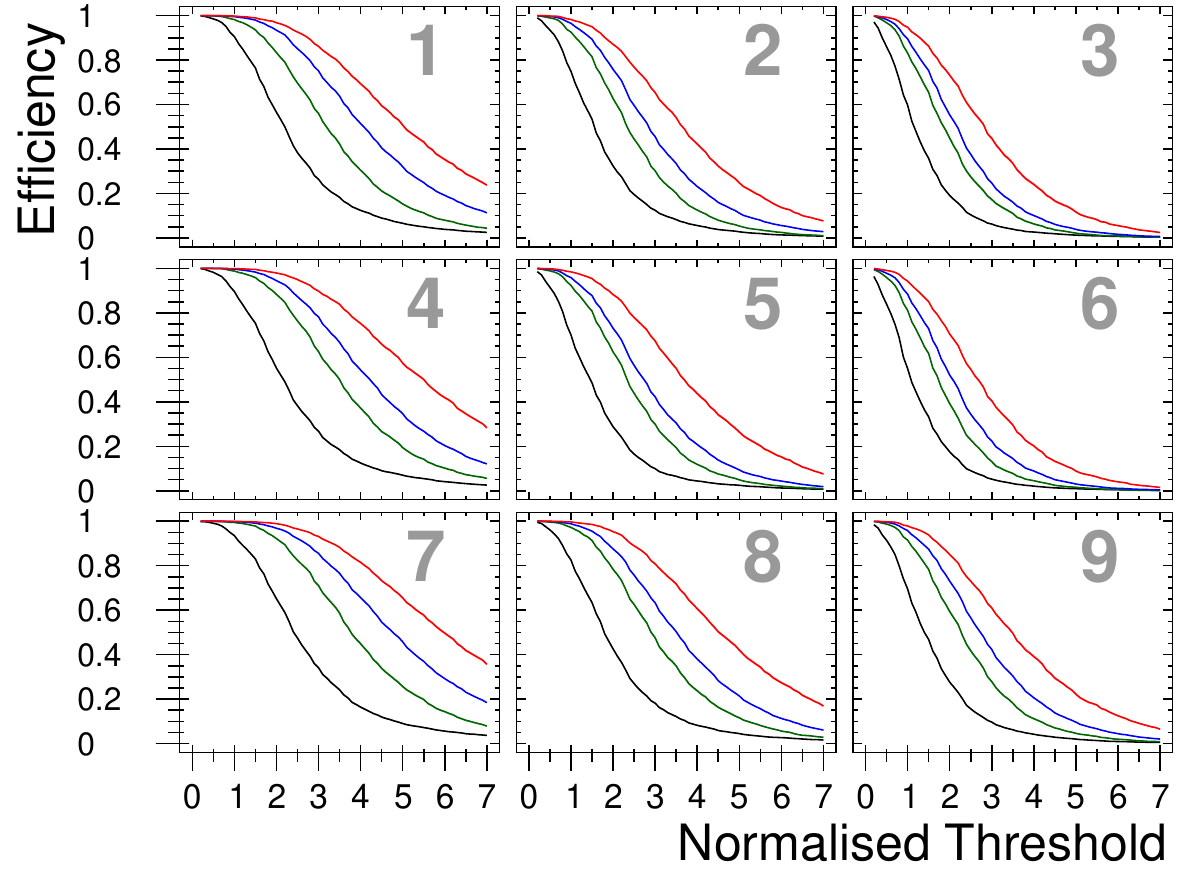}
        \subcaption{\Tz2 Counter: left side readout}
    \end{minipage}
    \hfill
    \begin{minipage}{0.47\textwidth}
        \centering
        \includegraphics[width=\textwidth]{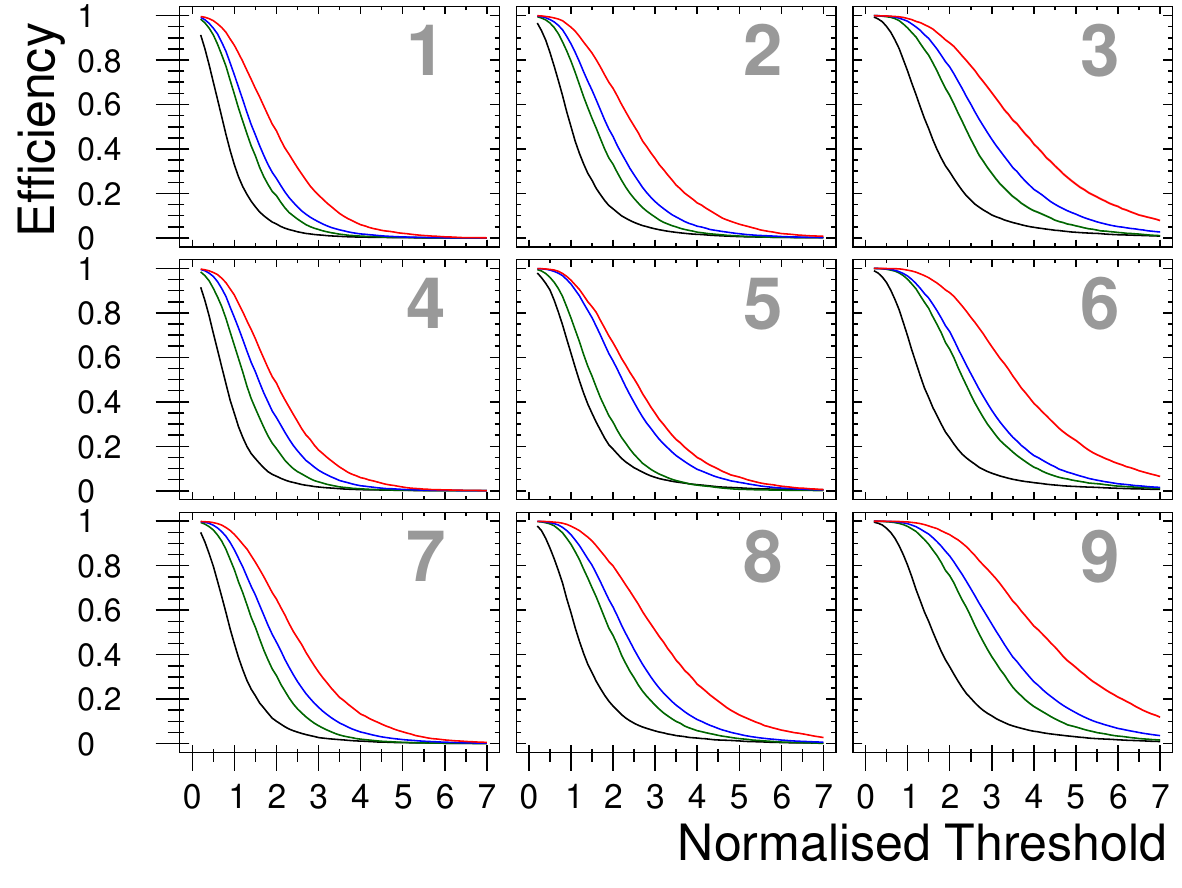}
        \subcaption{\Tz2 Counter: right side readout}
    \end{minipage}
    
    \vspace{0.02\textheight}
    
    \begin{minipage}{0.47\textwidth}
        \centering
        \includegraphics[width=\textwidth]{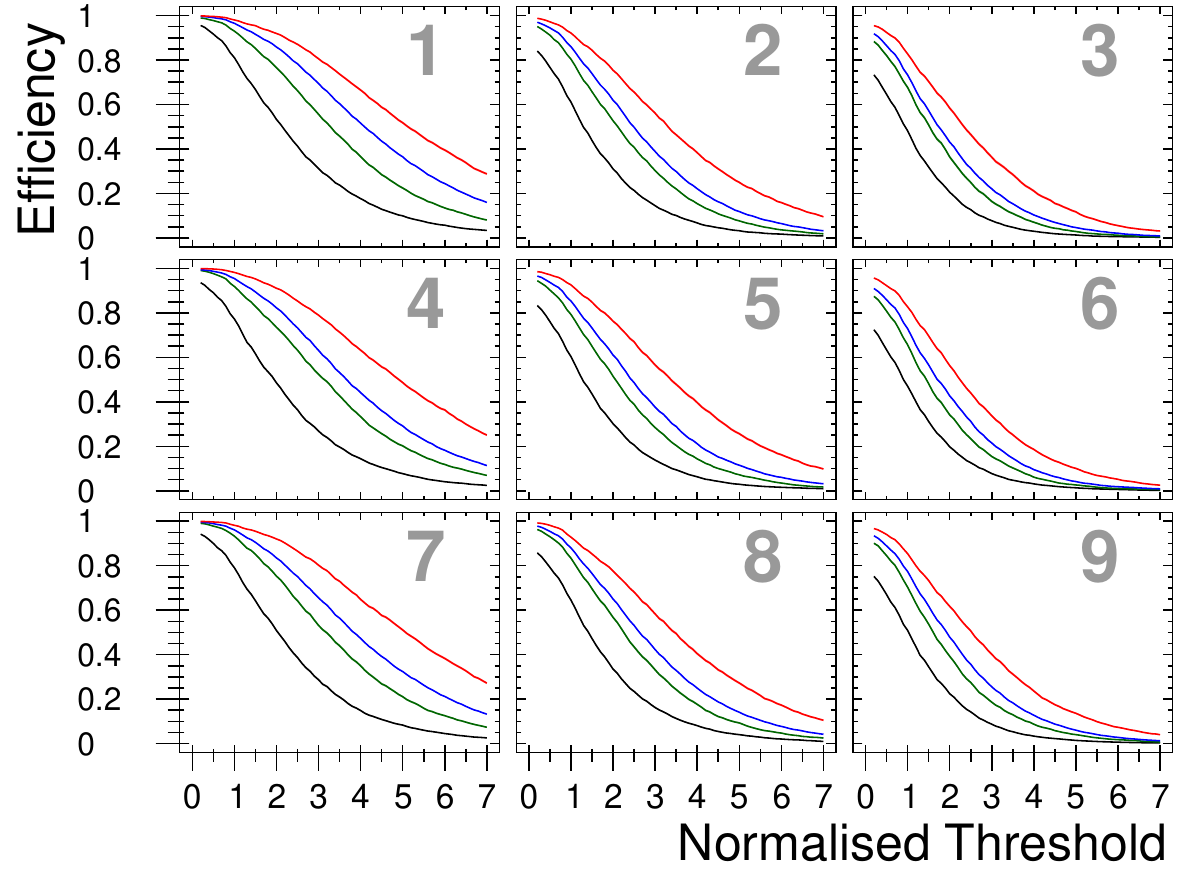}
        \subcaption{\Tz3 Counter: left side readout}
    \end{minipage}
    \hfill
    \begin{minipage}{0.47\textwidth}
        \centering
        \includegraphics[width=\textwidth]{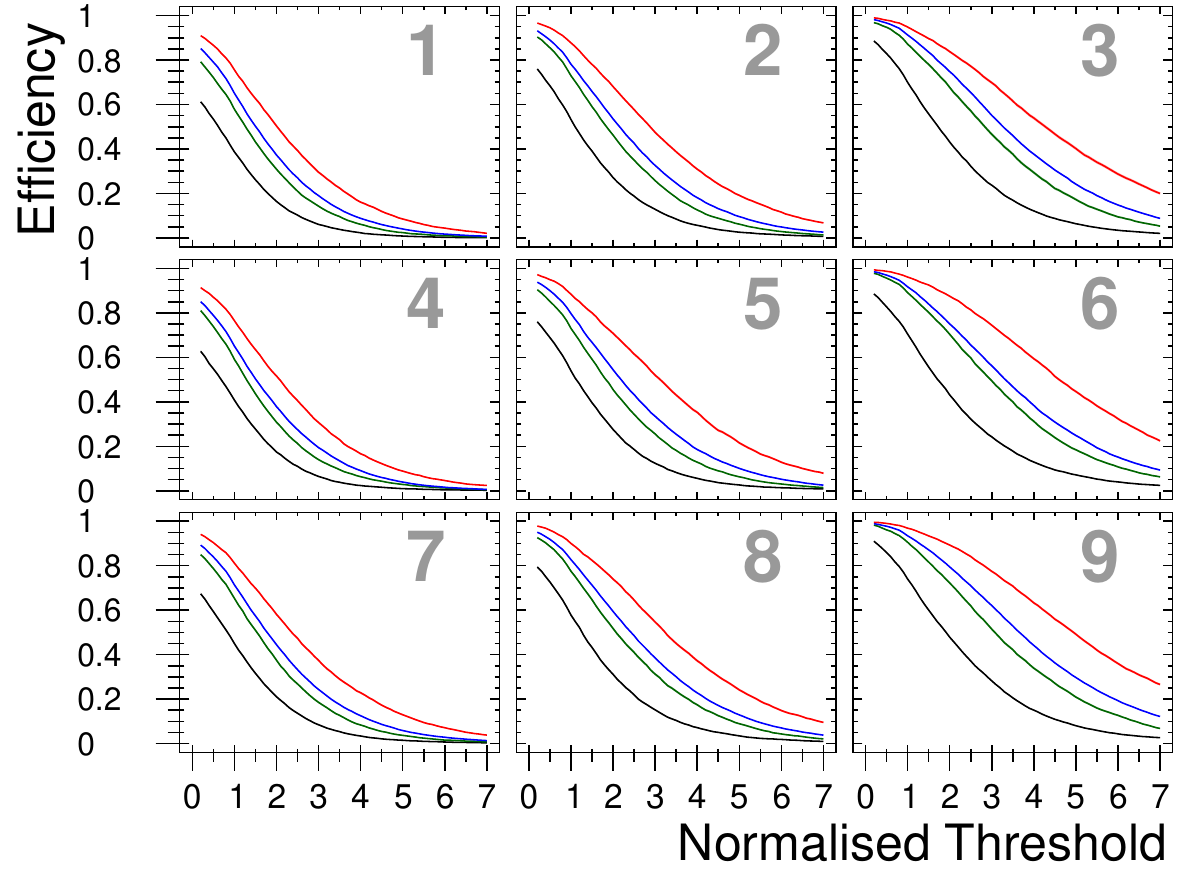}
        \subcaption{\Tz3 Counter: right side readout}
    \end{minipage}
    \caption{
        Muon-trigger efficiencies of the \Tz2 and \Tz3 Counters at the nine beam spots as functions of the waveform amplitude threshold.
        The ``left'' and ``right'' readout sides refer to waveforms from the left-side and right-side PMTs, respectively.
        The amplitude is normalised to the most probable value at the central beam spot for the 250~\MeVc{} muon beam.
        Line colours indicate degrader thicknesses: 0~mm (black), 48~mm (green), 50~mm (blue), and 54~mm (red).
        Statistical uncertainties are negligible.
    }
    \label{fig:EfficiencyCurves_BeamSpots}
\end{figure}

\fig{fig:EfficiencyCurves_BeamSpots} shows the evaluated \eff{trig} as functions of the normalised threshold amplitude, with variations across beam spots and degrader thicknesses.
As expected, the efficiency tends to decrease as the beam spot moves farther from the readout side.
As the muon momentum decreases, the energy deposit increases, leading to higher detection efficiency.  
Compared to the \Tz2 Counter, the \Tz3 Counter exhibits lower efficiency due to its larger cross section, which further suppresses light collection efficiency.  
While no significant differences are observed in the vertical direction due to structural symmetry, small variations are likely influenced by several factors in the manufacturing process, including the precision and uniformity of the bonding between the scintillator and light guide.  

\begin{figure}[h!]
    \centering
    \begin{minipage}{0.47\textwidth}
        \centering
        \includegraphics[width=\textwidth]{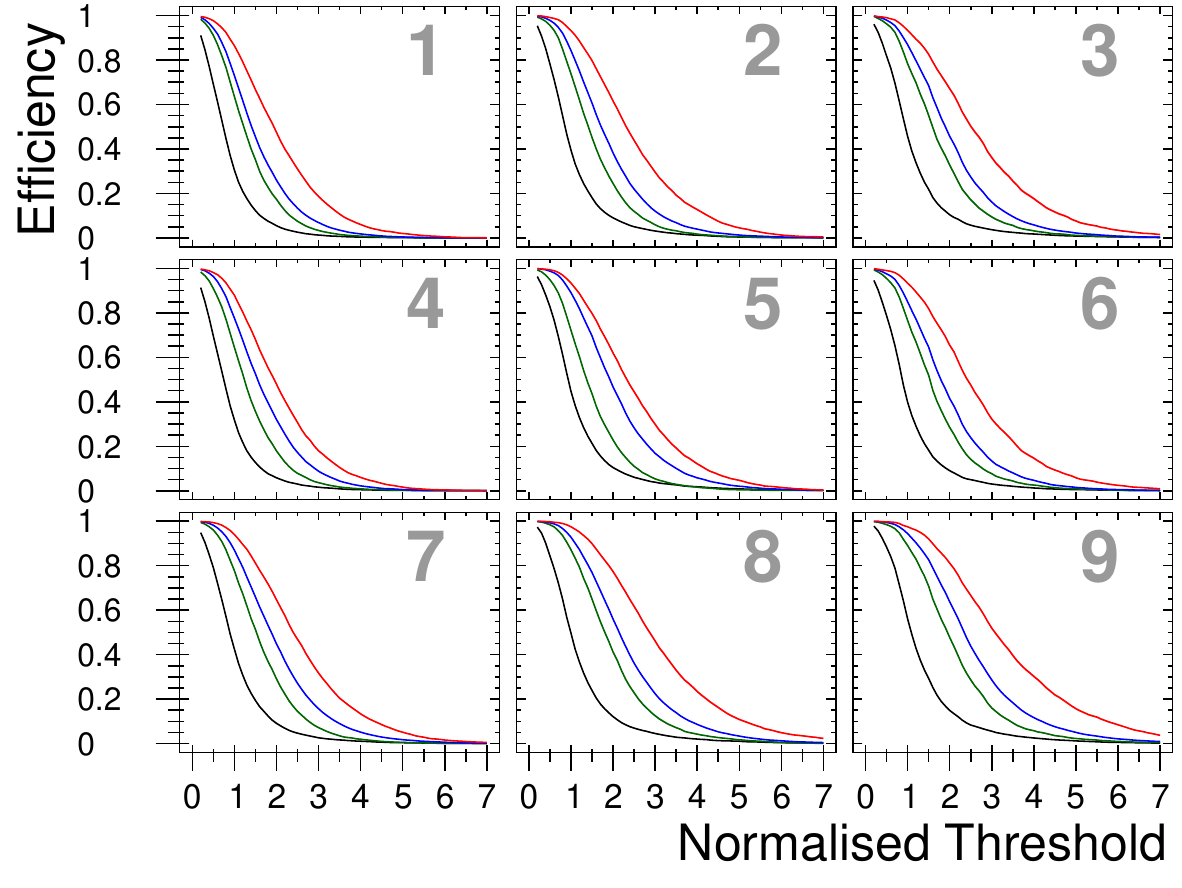}
        \subcaption{Coincidence approach: \Tz2 Counter}
    \end{minipage}
    \hfill
    \begin{minipage}{0.47\textwidth}
        \centering
        \includegraphics[width=\textwidth]{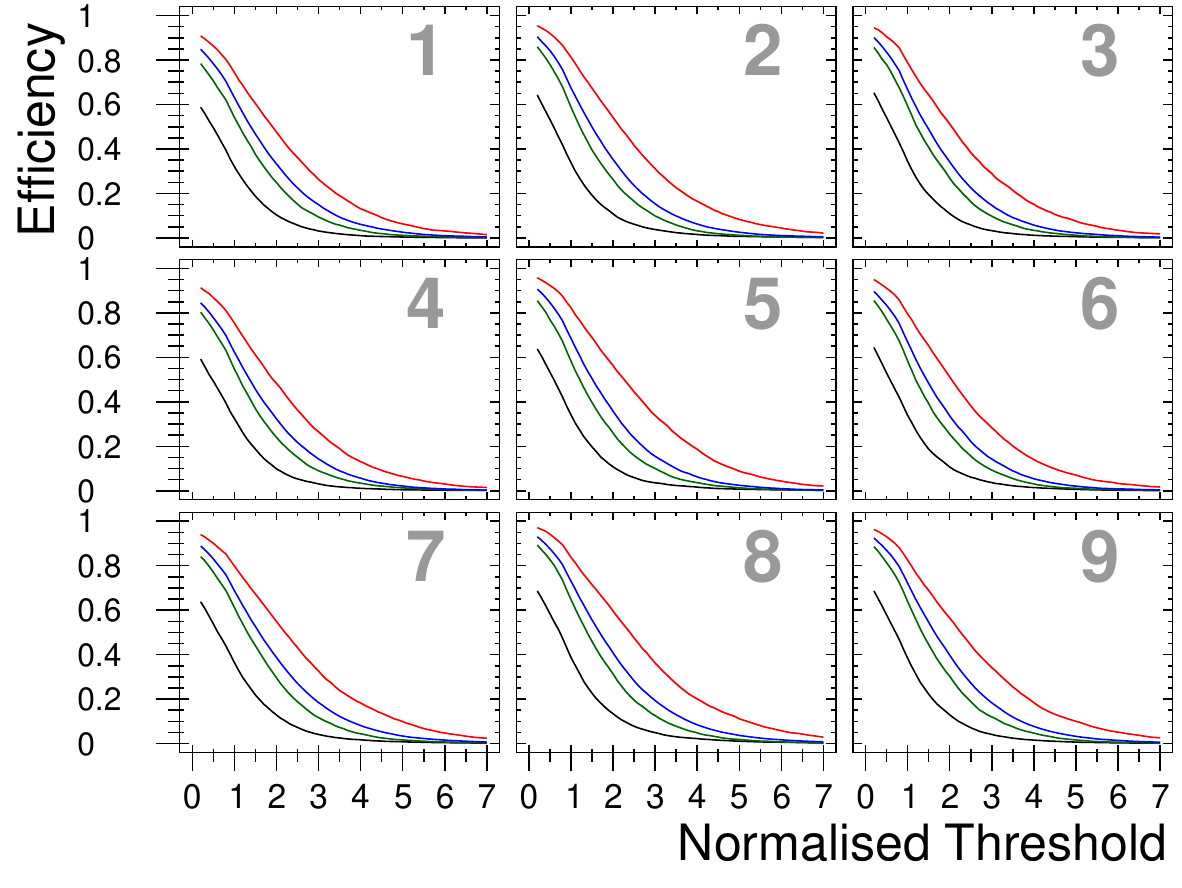}
        \subcaption{Coincidence approach: \Tz3 Counter}
    \end{minipage}

    \vspace{0.02\textheight}
    
    \begin{minipage}{0.47\textwidth}
        \centering
        \includegraphics[width=\textwidth]{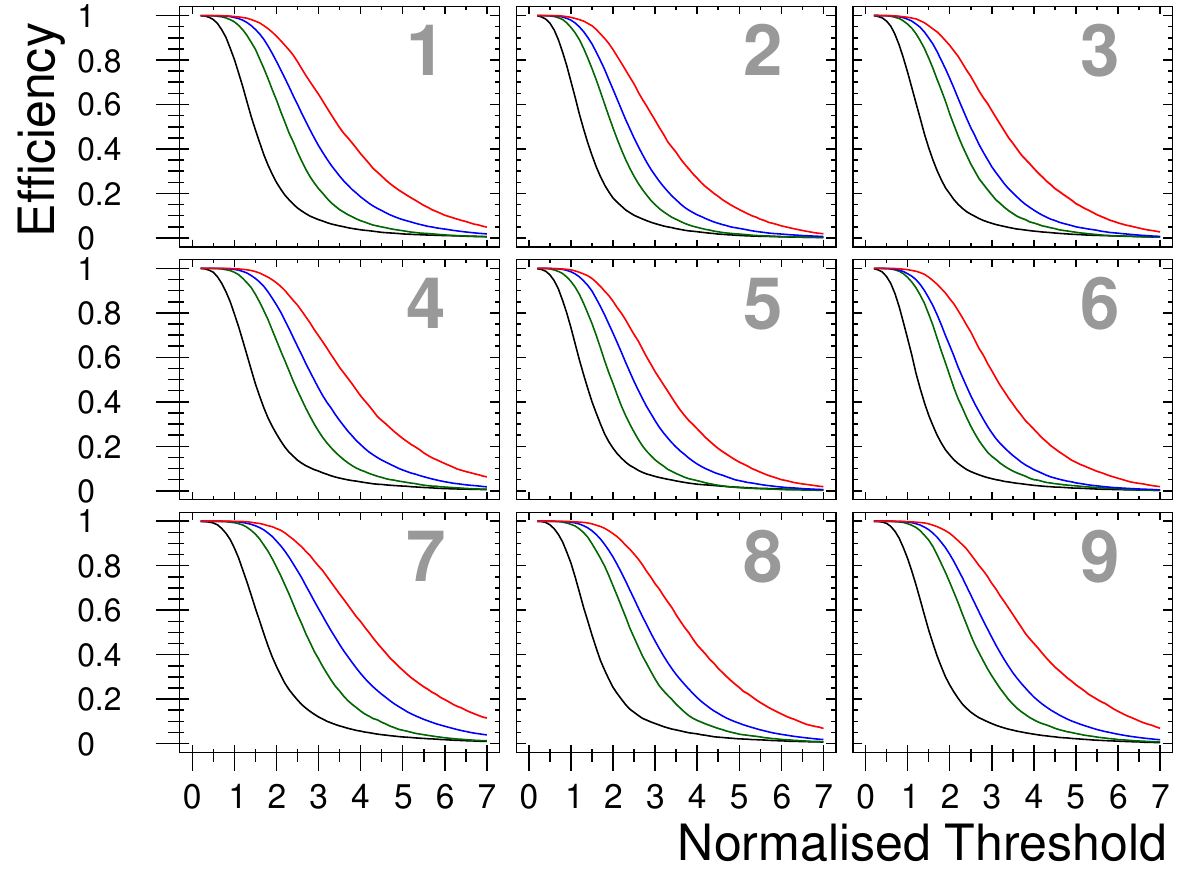}
        \subcaption{Waveform-summing approach: \Tz2 Counter}
    \end{minipage}
    \hfill
    \begin{minipage}{0.47\textwidth}
        \centering
        \includegraphics[width=\textwidth]{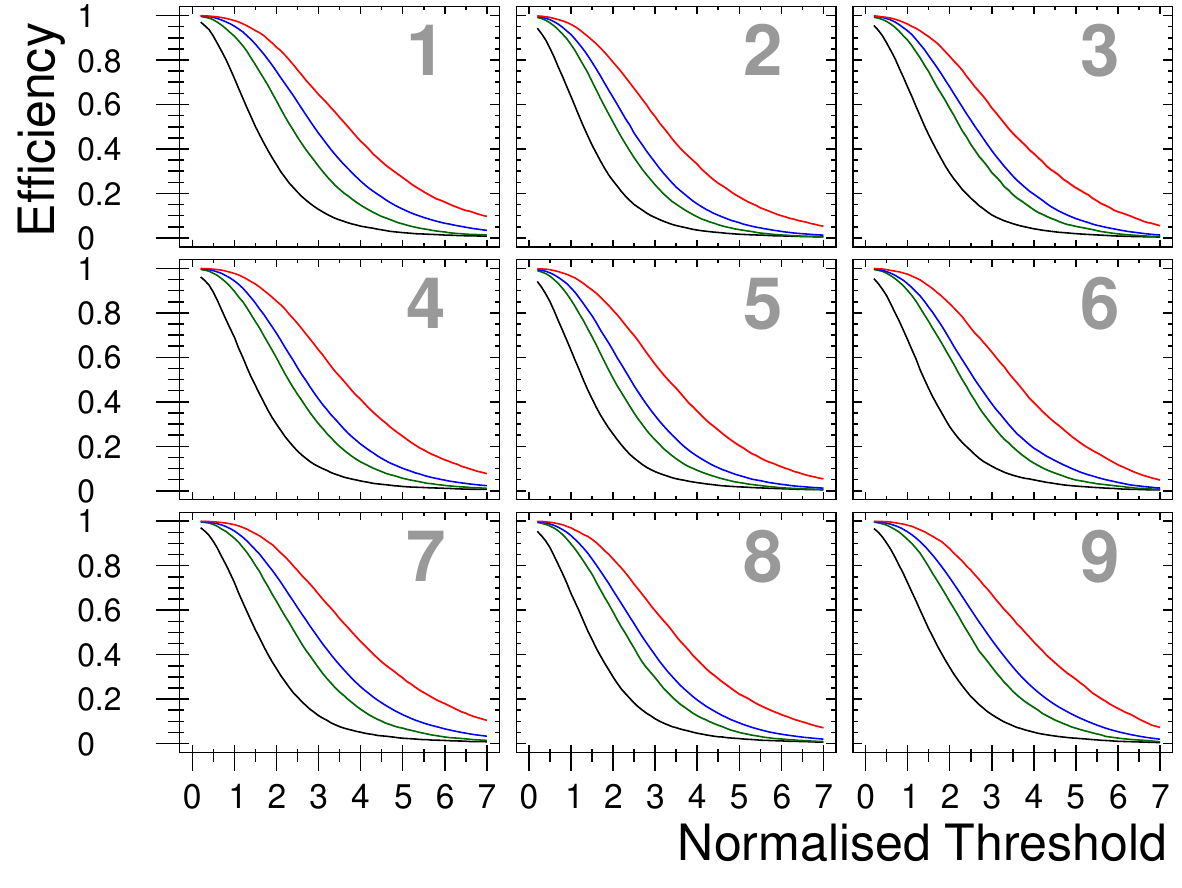}
        \subcaption{Waveform-summing approach: \Tz3 Counter}
    \end{minipage}
    \caption{
        Muon-trigger efficiencies of the \Tz2 and \Tz3 Counters as functions of the waveform amplitude threshold at nine beam spots, comparing two approaches for handling signals from their dual PMTs.
        (a, b) show the coincidence method, where both the left and right readout sides must simultaneously exceed the threshold.
        (c, d) show the waveform-summing method, where waveforms from both sides are summed before applying the threshold.
        The threshold is normalised to the most probable pulse peak amplitudes at the central beam spot for the 250~\MeVc{} muon beam in each case.
        Line colours indicate degrader thicknesses: 0mm (black), 48mm (green), 50mm (blue), and 54mm (red).
        Statistical uncertainties are negligible.
    }
    \label{fig:EfficiencyCurves_BeamSpots_Trig}
\end{figure}

% How to treat the Left and Right signals for trigger?
Based on the characteristics described above, we considered two methods for generating a trigger signal from the signals obtained at the left and right readout sides.
% COIN method
The first method is the coincidence approach, which accepts only those events in which both sides simultaneously exceed the individual thresholds.
\fig{fig:EfficiencyCurves_BeamSpots_Trig}(a,b) shows the resulting \eff{trig} values obtained using this logic, where, for simplicity, the same normalised threshold is applied to both readout sides.
Owing to insufficient photon propagation efficiency throughout the entire sensitive region, the overall performance is significantly compromised.
The advantage of this approach, however, is its robustness against accidental noise on either side, as it enforces coincidence in both readout sides.

% SUM method
The second method sums the waveforms from both readout sides before applying a single threshold.
The challenge with this method lies in the need to pre-calibrate the gain and timing between the two PMTs and maintain their relative stability throughout the data-taking period.
Nevertheless, since this method compensates for beam-spot dependence, it enables higher \eff{trig} and is therefore discussed in more detail below.

\fig{fig:EfficiencyCurves_BeamSpots_Trig}(c,d) presents the results using this approach.
Each waveform is first normalised to the MPV, as in the previous analysis, then averaged, and the efficiency is evaluated at each threshold.
This approach significantly reduces the beam-spot dependence, resulting in consistent behaviour across all spots.
Its overall performance also surpasses that of the coincidence method.

\begin{figure}[tb]
    \centering
    \begin{minipage}{0.48\columnwidth}
        \centering
        \includegraphics[width=\textwidth]{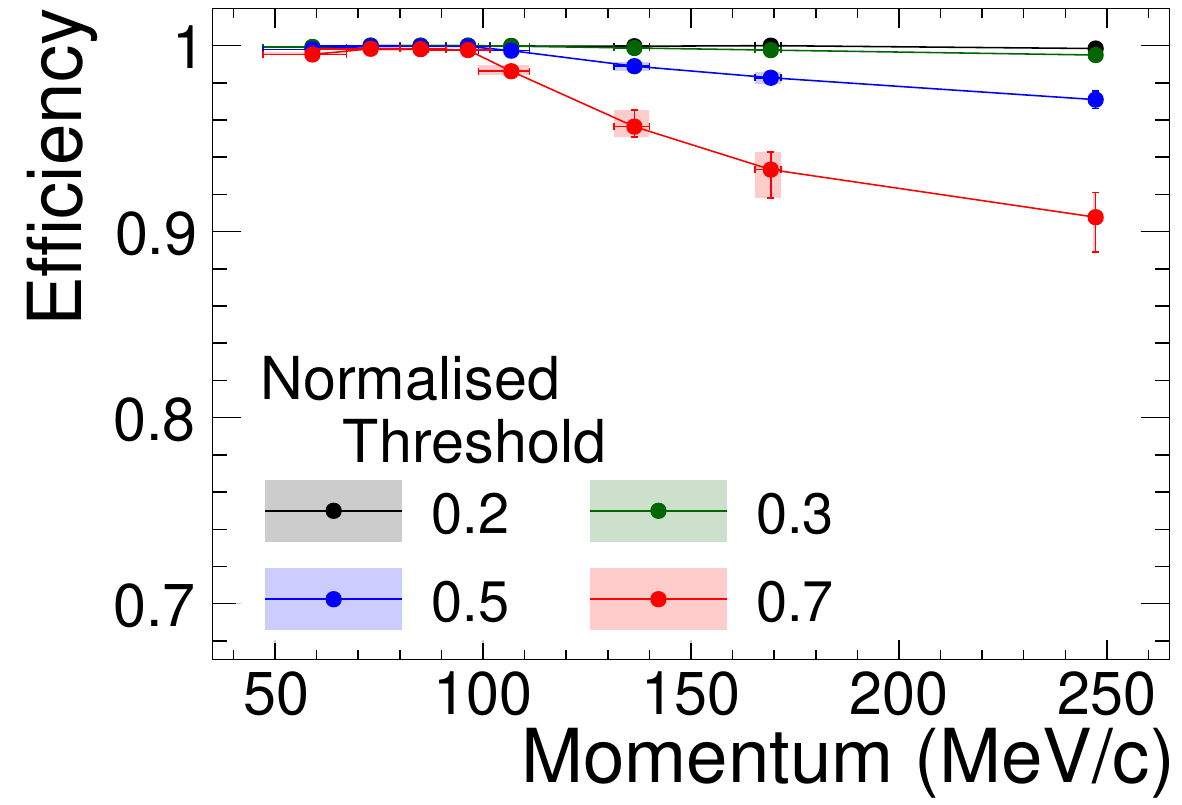}
        \subcaption{\Tz2 Counter}
    \end{minipage}
    \hfill    
    \begin{minipage}{0.48\columnwidth}
        \centering
        \includegraphics[width=\textwidth]{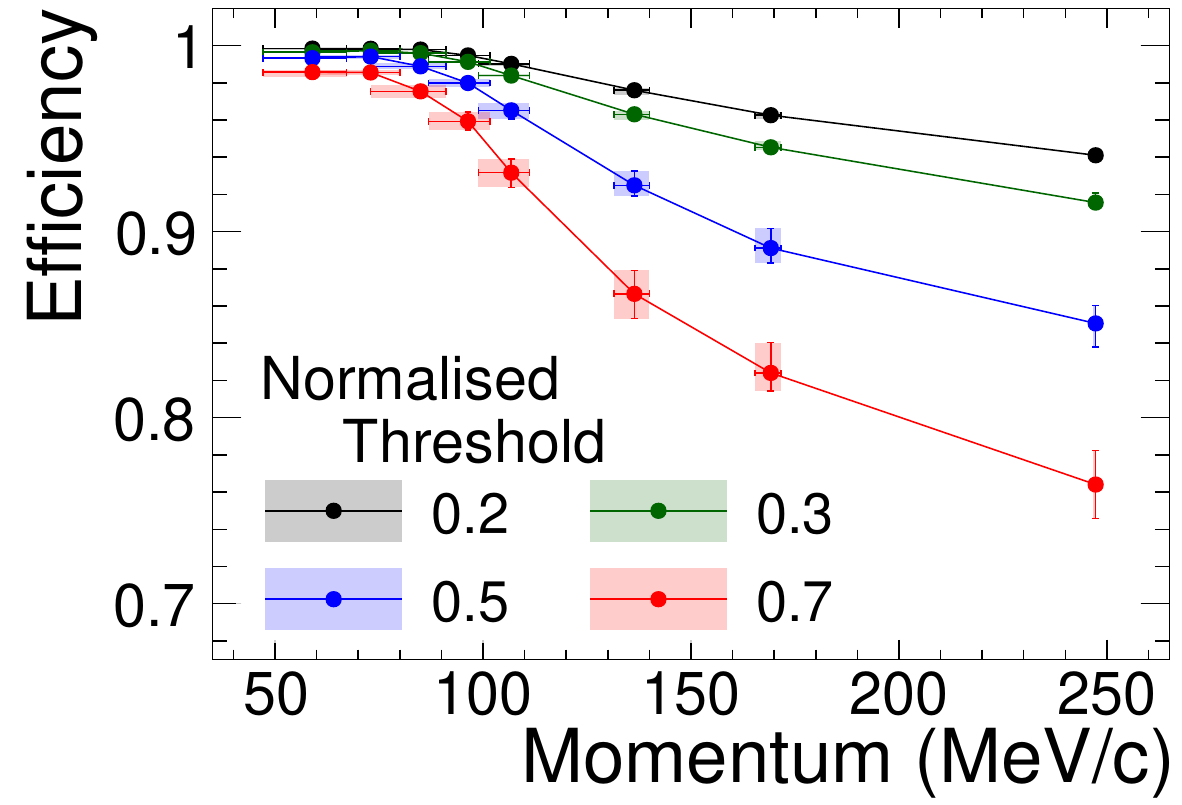}
        \subcaption{\Tz3 Counter}
    \end{minipage}
    \caption{
        Muon-trigger efficiencies of the \Tz2 and \Tz3 Counters at the central beam spots as functions of incident muon momentum.
        Threshold values in the legend correspond to those used on the horizontal axes of \fig{fig:EfficiencyCurves_BeamSpots_Trig}(c,d) with the waveform-summing approach.
        Uncertainties are primarily determined by systematic components.
    }
    \label{fig:EfficiencyCurves_Mom}
\end{figure}

\fig{fig:EfficiencyCurves_Mom} presents the same results at the central beam spot as functions of the muon momentum listed in \tab{tab:Momenta} with several thresholds.  
The vertical uncertainties are calculated from light-yield fluctuation due to PMT contact using optical grease and vinyl tape, which can be less stable than using optical cement.
To assess this potential instability, we repeated PMT attachment and detachment and measurement of light yield by irradiating the centre with collimated $\beta$-rays from a $\rm {}^{90}Sr$ source.
As a result, the relative fluctuations through repeated attachment and detachment were found to be $5.1\%$ and $5.8\%$ for the \Tz2 and \Tz3 Counters, respectively.

\begin{figure}[tb]
    \centering
    \begin{minipage}{0.32\columnwidth}
        \centering
        \includegraphics[width=\textwidth]{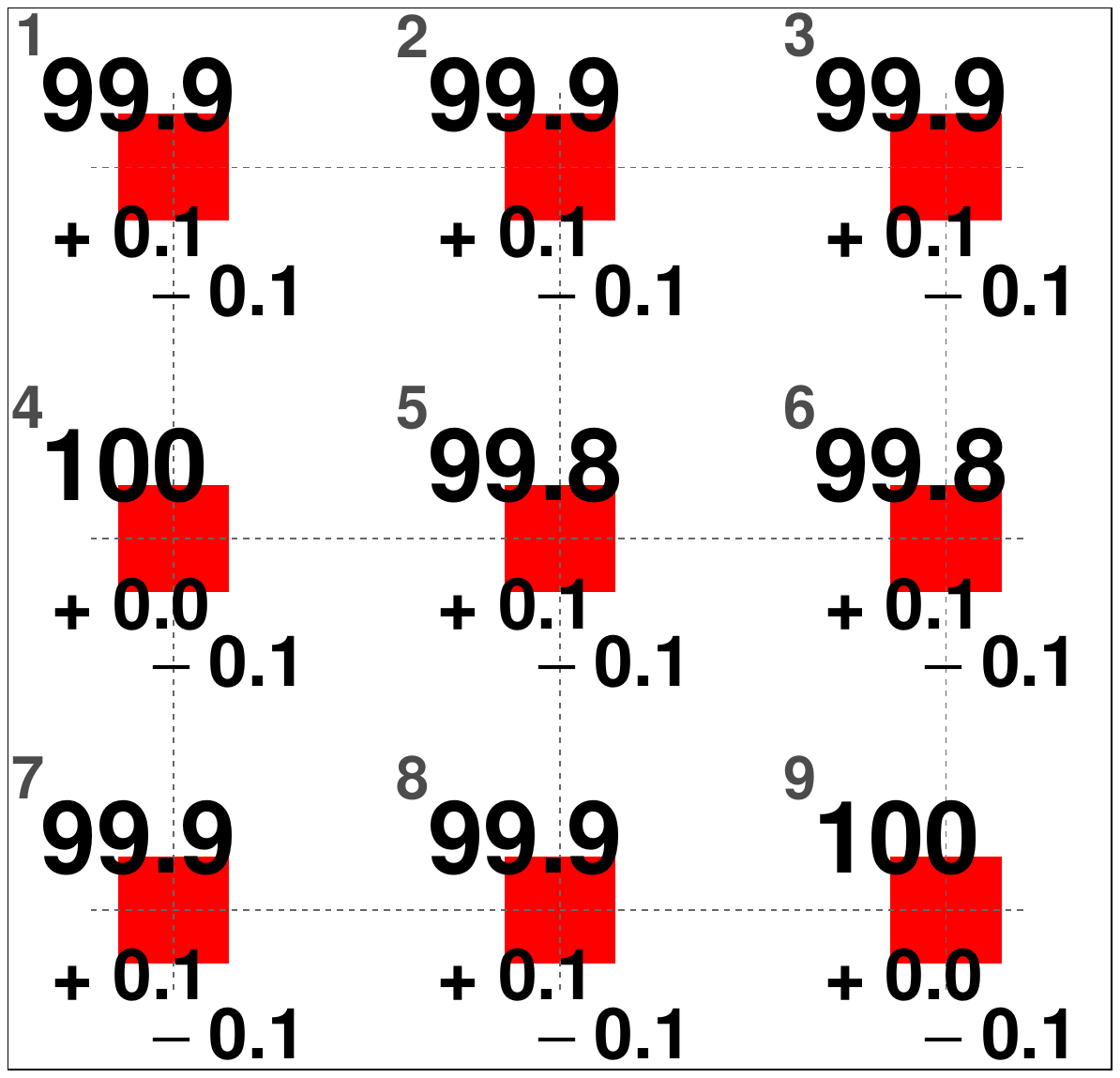}
        \subcaption{Threshold: 0.2}
    \end{minipage}
    \hfill
    \begin{minipage}{0.32\columnwidth}
        \centering
        \includegraphics[width=\textwidth]{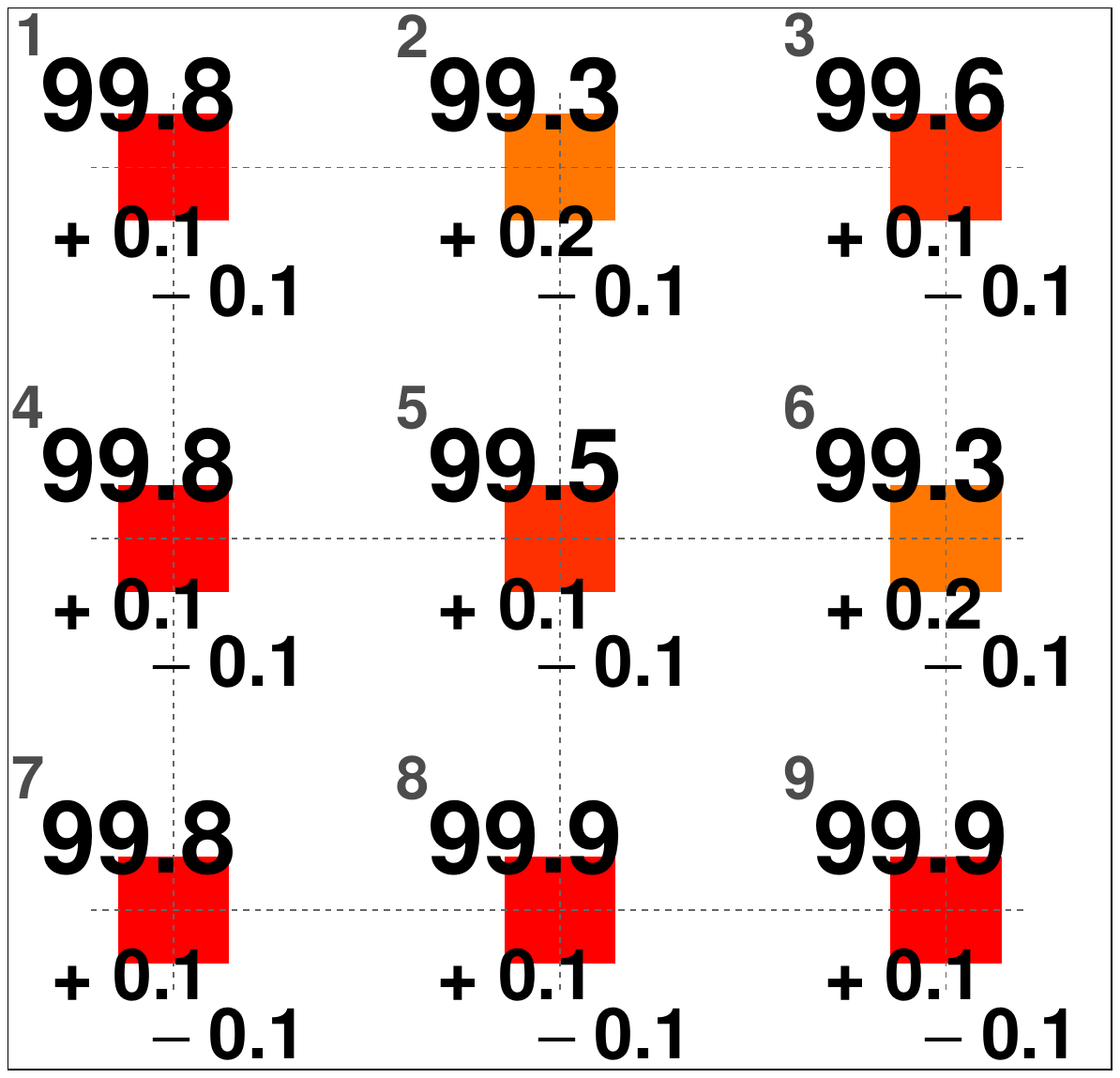}
        \subcaption{Threshold: 0.3}
    \end{minipage}
    \hfill                    
    \begin{minipage}{0.32\columnwidth}
        \centering
        \includegraphics[width=\textwidth]{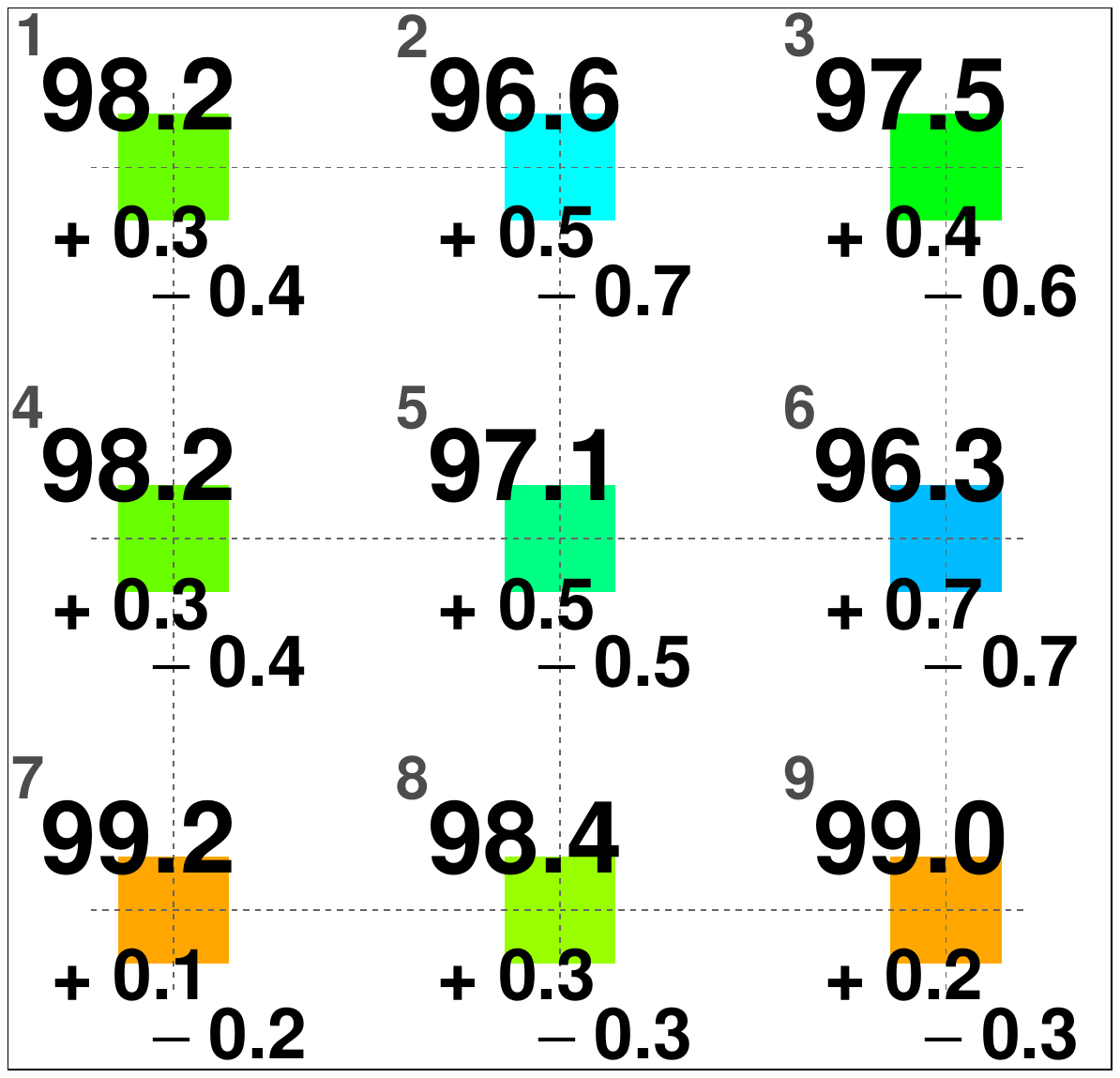}
        \subcaption{Threshold: 0.5}
    \end{minipage}
    \caption{
        Muon-trigger efficiencies (\%) of the \Tz2 Counter at the nine beam spots for a 250~\MeVc{} muon beam.
        The outlined frame indicates the boundary of the counter, while the filled areas and their colours represent the positions and efficiencies at each beam spot.
        Thresholds correspond to those on the horizontal axes of \fig{fig:EfficiencyCurves_BeamSpots_Trig}(c,d) with the waveform-summing approach.
        Uncertainties are primarily determined by systematic components.
    }
    \label{fig:EfficiencyMap}
\end{figure}

Furthermore, the systematic uncertainties of the \Tz2 Counter due to beam-spot dependence are evaluated.  
\fig{fig:EfficiencyMap} presents the \eff{trig} map at various thresholds for the most evident case of a 250~\MeVc{} muon beam with minimum-ionising energy deposits.
The dependence becomes pronounced at thresholds of $> 0.5$ that are unreasonably high in practical operation.  
This can be attributed to potential non-uniformity in optical collection efficiency introduced during production.

On the other hand, the dependence is significantly reduced for low-momentum muons of $< 100~\MeVc$ at realistic thresholds of $< 0.5$.
Under these conditions, \eff{trig} was evaluated with a minimum value of $99.97 \pm 0.01 \pm 0.03\%$, where the first uncertainty represents the average of all individual statistical uncertainties, and the second reflects the variation in mean values across the beam spots.

%%%%%%% Cosmic ray test
Although the waveform-summing method was found to effectively compensate for efficiency variations across the nine beam spots examined, we also evaluated the performance of the \Tz2 Counter at its extreme edges by conducting supplementary measurements with cosmic rays in the laboratory.  
Three specific positions were investigated:  
(2') above beam spot~2,  
(3') the upper corner diagonally from beam spot~3, and  
(6') to the right of beam spot~6.  
At each location, cosmic-ray events were selected by placing the \Tz2 Counter between the TRG1 and TRG2 counters, centred 15~mm inward from the counter’s edge or corner.

However, it is difficult to directly compare these measurements with those from the beam test.  
Compared to the 250~\MeVc{} muons used in the beam test, cosmic rays typically deposit 30--50\% less energy in the \Tz2 Counter on average.  
As a result, the deposited energy may be insufficient to produce even a single photoelectron in one of the PMTs, and the efficiency does not reach 100\% even at a near-zero threshold.  
To evaluate relative differences under these conditions, beam spot~5 was also re-measured using cosmic rays and employed as a reference.  
Waveform amplitudes and thresholds are normalised to the MPV of the pulse peak amplitude at spot~5, and the summing approach is applied.  
All the spots are expected to exhibit similar efficiencies at a practical threshold unless the light yield or propagation is locally degraded.

The efficiencies evaluated for each spot at a threshold of~0.2 are: spot~5, $98.0^{+0.6}_{-0.7}\%$; spot~2', $96.2^{+0.9}_{-1.1}\%$; spot~3', $96.8^{+0.8}_{-1.0}\%$; and spot~6', $99.2^{+0.4}_{-0.8}\%$, where the uncertainties represent statistical errors only.  
Only small differences are observed relative to the reference (spot~5).  
Given that systematic variations due to detector fabrication should additionally exist, as discussed in \fig{fig:EfficiencyMap}, we do not expect any significant performance degradation even when the muon momentum decreases to our target range.  
We conclude that the performance across the entire area of the \Tz2 Counter can be well preserved using the waveform-summing approach.

%%%%%%%%%%%%%%%%%%%%%%%%%%%%%%%%%%%%%%%%%%%%%%%%%%%%%%%%%%%%%%%%%%%%%%%%%%%%%%%%%%%%%%%%%%%%
% Electron-detection Efficiency
%%%%%%%%%%%%%%%%%%%%%%%%%%%%%%%%%%%%%%%%%%%%%%%%%%%%%%%%%%%%%%%%%%%%%%%%%%%%%%%%%%%%%%%%%%%%
\subsection{Electron-detection efficiency}

\eff{DIO} was evaluated for each of the \T1 and \T2 Counters using a 250~\MeVc{} electron beam.
In contrast to the \eff{trig} evaluation, which assumes the \PhaseAlpha{} trigger scheme with simple online signal processing, detailed waveform analysis was applied to identify pulses in the recorded waveforms.
The analysis integrates the waveform over the signal pulses and counts events where the integrated charge exceeds a specified threshold.

\begin{figure}[htb]
    \centering
    \begin{minipage}{0.48\columnwidth}
        \centering
        \includegraphics[width=\textwidth]{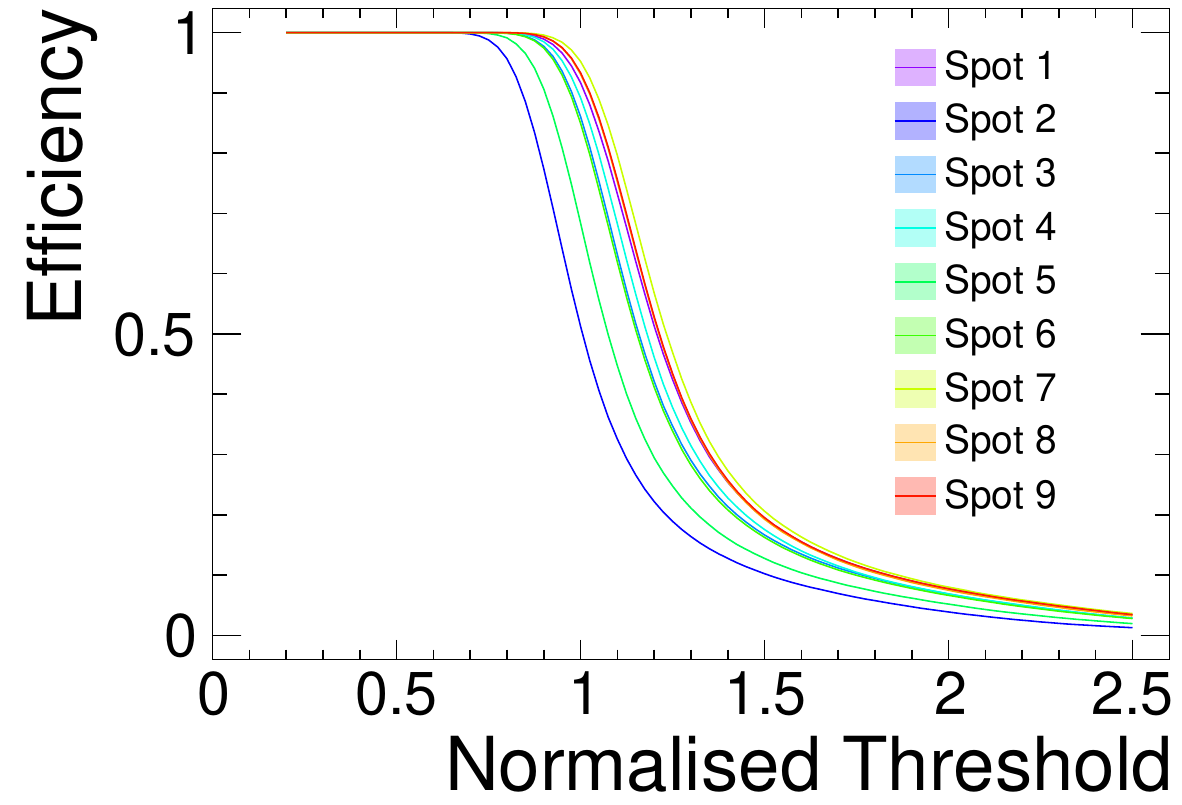}
        \subcaption{\T1 Counter}
    \end{minipage}
    \hfill
    \begin{minipage}{0.48\columnwidth}
        \centering
        \includegraphics[width=\textwidth]{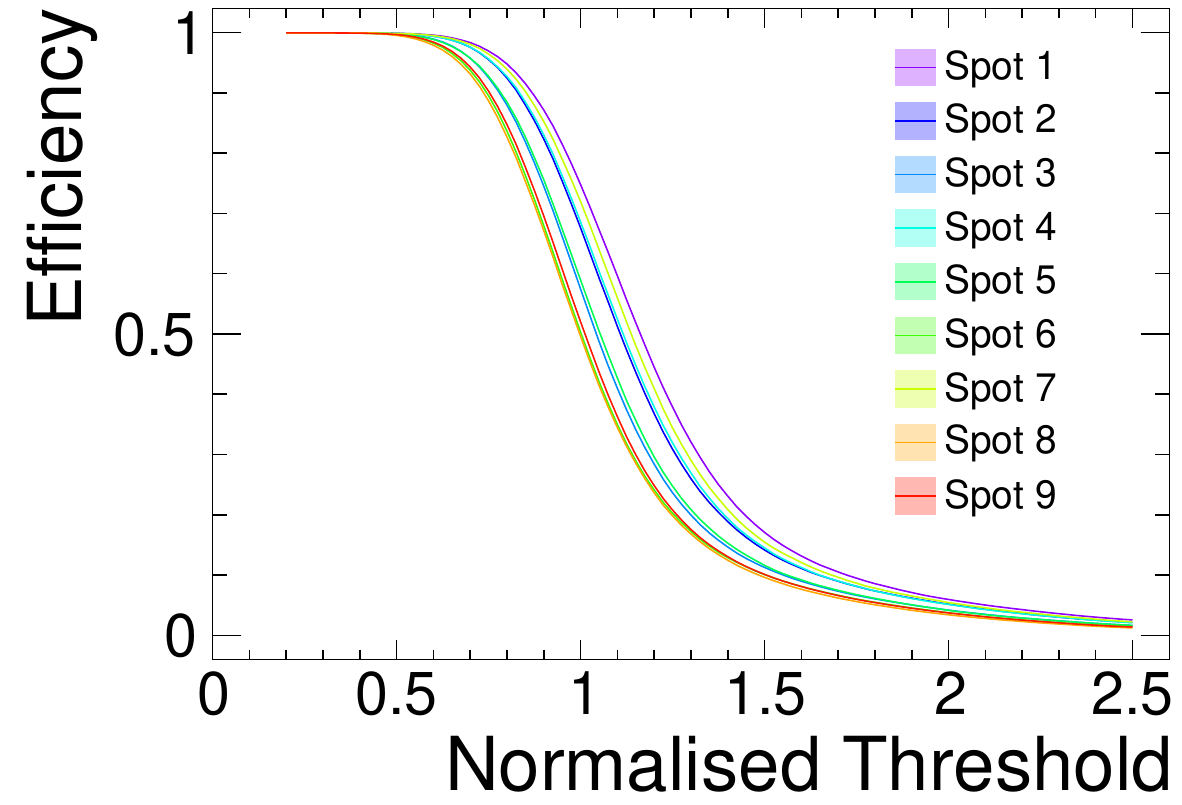}
        \subcaption{\T2 Counter}
    \end{minipage}
    \caption{
        Electron-detection efficiencies of the \T1 and \T2 Counters as functions of the threshold at nine beam spots.
        Thresholds are normalised to the most probable energy deposit at the central beam spot for each counter.
    }
    \label{fig:DetectionEfficiency}
\end{figure}

\fig{fig:DetectionEfficiency} shows the results as functions of the charge threshold at the nine beam spots.
The threshold is normalised to the MPV of the charge distribution at the central beam spot for each counter.
The curves start to drop around a threshold of 0.7 for the \T1 Counter and around 0.5 for the \T2 Counter, reflecting the thinner thickness of the latter.
The dropping points vary depending on the beam spot: beam spots closer to the corresponding readout side show relatively better performance, although individual variations are not negligible due to certain non-uniformities introduced during counter production.

At the displayed minimum threshold of 0.2, which is sufficiently above the pedestal noise amplitude, the efficiencies are high enough, resulting in $(99.999^{+0.001}_{-0.002} \pm 0.001)\%$ for the \T1 Counter and $(99.949^{+0.005}_{-0.006} \pm 0.007)\%$ for the \T2 Counter.
The first uncertainty represents the average of the individual statistical uncertainties, and the second reflects the variation across the beam spots.  
Even at higher thresholds, the efficiencies remain $(99.97 \pm 0.07)\%$ at 0.7 for the \T1 Counter and $(99.75 \pm 0.11)\%$ at 0.5 for the \T2 Counter, with the uncertainties dominated by the second term.

Therefore, it is concluded that \eff{DIO} and its fluctuations are negligibly small around the analysis cuts for \pur{DIO} and \accep{DIO}.
Furthermore, since the minimum-ionising electrons used in this evaluation deposit approximately 30\% less energy in MPV than DIO electrons, the results provide a conservative validation.

%%%%%%%%%%%%%%%%%%%%%%%%%%%%%%%%%%%%%%%%%%%%%%%%%%%%%%%%%%%%%
% Conclusion
%%%%%%%%%%%%%%%%%%%%%%%%%%%%%%%%%%%%%%%%%%%%%%%%%%%%%%%%%%%%%
\section{Conclusion}
\label{sec:Conclusion}

The Range Counter (RC) developed for the COMET \PhaseAlpha{} experiment has been designed, constructed, and evaluated to measure the momentum spectrum of transported muons.  
It exploits the characteristic lifetime of muonic atoms in a heavy material to count the number of muons entering the RC.  
Systematic performance assessments confirm its capability to reconstruct the number of stopped muons in the absorber.

The absorber configuration was optimised through comparison between copper and aluminium, and copper---having a decay-in-orbit (DIO) rate of $7.20\pm0.04\%$---was selected due to its shorter muonic atom lifetime, which allows for clearer separation between signal and background events.

For the \T0 Counter, a 0.5~mm thick, 20~cm square plastic scintillator plate was adopted based on an experimental performance comparison with a 30~cm square type.  
This choice represents an optimal balance between mechanical strength and light collection efficiency, while preserving the thinness required for RC performance.  
Experimental studies confirmed that, although the muon-trigger efficiency varies across the counter due to limited light-collecting capability, summing the waveforms from the dual PMTs effectively reduces this dependence.  
A stable performance of $\eff{trig} = 99.97\pm 0.01\pm 0.03\%$ was achieved over the entire counter area.

The detection efficiencies of the \T1 and \T2 Counters were also experimentally evaluated with minimum-ionising electrons.  
The analysis results show a stably high performance of $> 99\%$ with a reasonably high threshold.

The purity and acceptance of the \T1 and \T2 Counters for reconstructing the number of detected DIO electrons were validated through simulation.  
In particular, the purity strongly depends on the energy deposit threshold used to cut muon capture-induced background events.  
At a maximum pragmatic threshold, the purity and acceptance are estimated to be roughly 60\% and 47\%, respectively.

Overall, the RC has been demonstrated to be a simple and high-performance detector for negative muon beam measurements.  
The methodologies developed for its design and evaluation provide a solid foundation for muon beam diagnostics in \PhaseAlpha{} and offer valuable insights for future experiments requiring slow-muon measurements.

%%%%%%%%%%%%%%%%%%%%%%%%%%%%%%%%%%%%%%%%%%%%%%%%%%%%%%%%%%%%%
% Acknowledgements
%%%%%%%%%%%%%%%%%%%%%%%%%%%%%%%%%%%%%%%%%%%%%%%%%%%%%%%%%%%%%
\section*{Acknowledgements}
The authors would like to express their gratitude to the members of the COMET collaboration who participated in the detector evaluation experiments.
We also appreciate the technical staff of Imperial College London for their valuable support.
The experiments were conducted at the Paul Scherrer Institut in Switzerland and the Materials and Life Science Experimental Facility at J-PARC, Japan, under Proposal No. 2022A0169.
We are grateful to the local support staff at both facilities for their assistance.

These studies were supported and funded by the European Union’s Horizon 2020 Marie Skłodowska-Curie Actions under Grant Agreement Number 101033237, the European Union’s Horizon Europe Research and Innovation programme under Grant Agreement Number 101057511 (EURO-LABS), and JSPS KAKENHI Grant Numbers 21H04971 and 22H00139.

%%%%%%%%%%%%%%%%%%%%%%%%%%%%%%%%%%%%%%%%%%%%%%%%%%%%%%%%%%%%%
% Bibliography
%%%%%%%%%%%%%%%%%%%%%%%%%%%%%%%%%%%%%%%%%%%%%%%%%%%%%%%%%%%%%
\bibliographystyle{elsarticle-num} 
\bibliography{biblio}

\end{document}